# The *Planck* mission

François R. Bouchet

Institut d'Astrophysique de Paris (IAP), CNRS &
UPMC, 98 bis Bd Arago, F75014, Paris

On behalf of the *Planck* collaboration for the results. All further mistakes are mine :-)

These lecture notes discuss some aspects of the *Planck*[1] mission, whose prime objective was a very accurate measurement of the temperature anisotropies of the Cosmic Microwave Background (CMB). We announced our findings a few months ago, on March 21$^{st}$, 2013. I describe some of the relevant steps we took to obtain these results, sketching the measurement process, how we processed the data to obtain full sky maps at 9 different frequencies, and how we extracted the CMB temperature anisotropies map and angular power spectrum. I conclude by describing some of the main cosmological implications of the statistical characteristics of the CMB we found. Of course, this is a very much shortened and somewhat biased view of the *Planck* 2013 results, written with the hope that it may lead some of the students to consult the original papers.

## 0.1 From wishes and hopes to bits on the ground

Most of the background photons are actually Cosmic Microwave Background (CMB) photons, as can be visualised on Fig. 0.1. Their spectral distribution is accurately fit by a Planck distribution with a temperature of 2.7255±0.0006 K (Fixsen, 2009), with therefore a number density of about 410 photons per cubic centimetre. This shape was a definite prediction of the so-called "Big Bang" model with early predictions of the expected temperature today of about 5 Kelvin (Alpher and Herman, 1948). In this expanding model, which we adopt as a framework throughout these lectures, this "Planckian" shape is acquired very early during a hot and dense phase, at a redshift $z \gtrsim 10^7$, when reactions which do not conserve the photon number (bremsstrahlung, $e \to e\gamma$, double-Compton $e \to e\gamma\gamma$) froze out, i.e. when their characteristic times became larger than the expansion time ($H^{-1} = (\dot a/a)^{-1}$, if $a$ stands for the scale factor of the metric). Later on, the photon distribution function kept constant but for a temperature decrease proportional to the expansion factor $a$.

In the early Universe, baryons and photons were tightly coupled through Thomson scatterings of photons by free electrons (and nuclei equilibrate collisionally with electrons). When the temperature in the Universe becomes smaller than about 3000 K (which is much lower than 13.6 eV due to the large number of photons per baryons $\sim 1.5 \cdot 10^9$), the cosmic plasma recombines and the ionisation rate, $x_e$, falls from unity at $z > 1100$ down to $x_e < 10^{-3}$ at

---

[1] *Planck* (http://www.esa.int/Planck) is a project of the European Space Agency (ESA) with instruments provided by two scientific consortia funded by ESA member states (in particular the lead countries France and Italy), with contributions from NASA (USA) and telescope reflectors provided by a collaboration between ESA and a scientific consortium led and funded by Denmark.

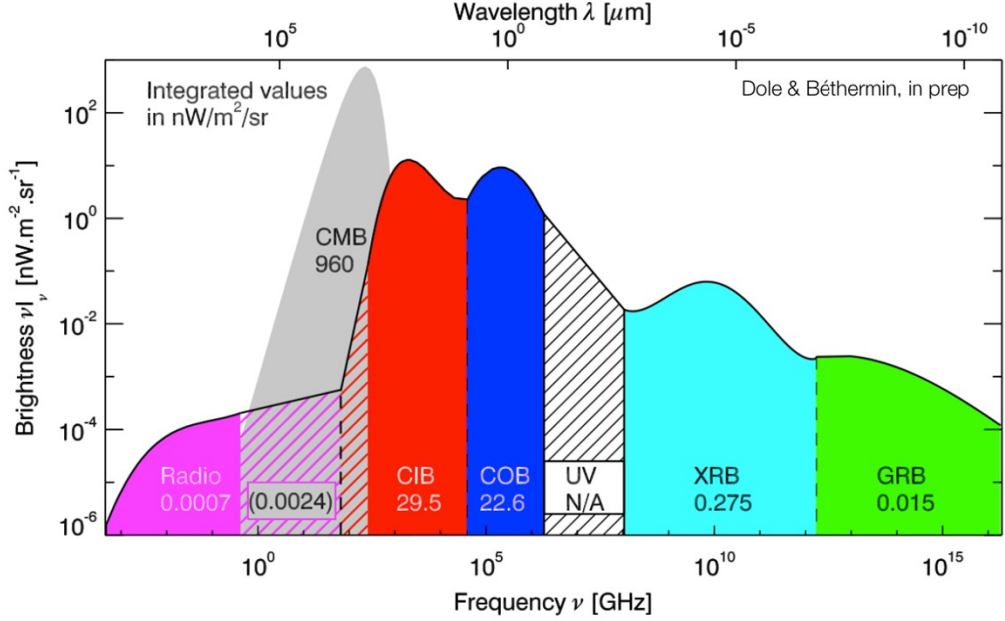

**Fig. 0.1** Cosmic background frequency spectrum, from radio to gamma rays, after Dole & Bétermin. The intensity is shown per logarithmic frequency interval, so that the energy within different bands can be directly compared.

$z < 1100$: the photons mean free path $\propto 1/x_e$ rapidly becomes much larger than the Horizon $\sim cH^{-1}$. As a result, the Universe becomes transparent to background photons, over a narrow redshift range of 200 or less. Photons will then propagate freely as long as galaxies and quasars do not reionise the Universe (but by then the electron density will have fallen enough that only a small fraction will be re-scattered). We therefore observe a thin shell around us, the last scattering "surface" where the overwhelming majority of photons last interacted (excluding gravitationally) with baryonic matter at a redshift of 1100, when the Universe was about 380 000 years old.

In the standard cosmological model, some physical process in the very early Universe generates the seed fluctuations which give rise to all the large scale structures (hereafter LSS) we see today through their development by gravitational instability. The evolution of primordial fluctuations can be accurately followed, and it was long ago predicted that, in order to account for the formation of large scale structure, their imprint as temperature fluctuations should have an *rms* of $\sim 100\,\mu K$ in the presence of cold dark matter (CDM). The smallness of these fluctuations is indeed why it took so long to first detect them.

To analyse the statistical properties of the temperature anisotropies, we can either compute the angular correlation function of the temperature contrast, $\delta_T$, or the angular power spectrum $C(\mathbf{f})$ which is it's spherical harmonics transform (in practice, one transforms the $\delta_T$ pattern in harmonic modes, $a_{\mathbf{f},m}$, and sums over the $m$'s at each multipole since the pattern should be isotropic – at least for the trivial topology). A given multipole corresponds to an angular scale $\theta \sim 180°/\mathbf{f}$. How this is measured in practice is discussed in §0.3.2. These two-point statistics characterise completely a Gaussian field. As will be reviewed in other contributions of this volume, detailed predictions have been made from the seventies onward, *preceding observations*, and providing a well-developed framework to think the next steps, both theoretically and observationally. An excellent introduction to the physics of the CMB anisotropies can be found in the book by V. Mukhanov "Physical Foundations of Cosmology" (Mukhanov, 2005), which is available on line at http://ebooks.cambridge.org/ebook.jsf?bid=CBO9780511790553. Equally excellent is the book by P. Peter and J.-P. Uzan "Primordial Cosmology" (Peter and Uzan, 2009), with the additional merit - for some - of an edition in French.

The first clear detection of the CMB anisotropies was made in 1992 (Smoot *et al.*, 1992; Wright *et al.*, 1992) by the DMR instrument aboard the *COBE* satellite orbiting the earth (and soon afterwards by the FIRS experiment), with a ten degree (effective) beam and a signal to noise per resolution element around unity (note though that the dipolar pattern had been detected earlier). This lead to a clear detection of the large scale, low-$\ell$, Sachs-Wolf effect, at low multipole (see fig. 0.2.a) indicating that the logarithmic slope of the primordial power spectrum, $n_s$, could not be far from one. The ~ 30 µK height of the low-$\ell$ plateau detected by DMR gave a direct estimate of the normalisation of the spectrum, $A_s$ (assuming the simplest theoretical framework, without much possible direct checks of the other existing theoretical predictions, given the data). This was quickly followed by many other results, and Fig. 0.2.b shows that, by 1999, one could already "see" the first acoustic peak of the spectrum. With time, many other features of the inflationary cosmology were progressively unveiled, by a very large number of ground and balloon experiments, and by the second generation space experiment WMAP, whose final results based on 9 years of operations were released at the time of *Planck* first cosmological announcement (Bennett *et al.*, 2012a; Hinshaw *et al.*, 2012). A detailed account of pre-*Planck* CMB experiments can be found in particular in the book "Finding the Big Bang" (Peebles, Page and Partridge, 2009).

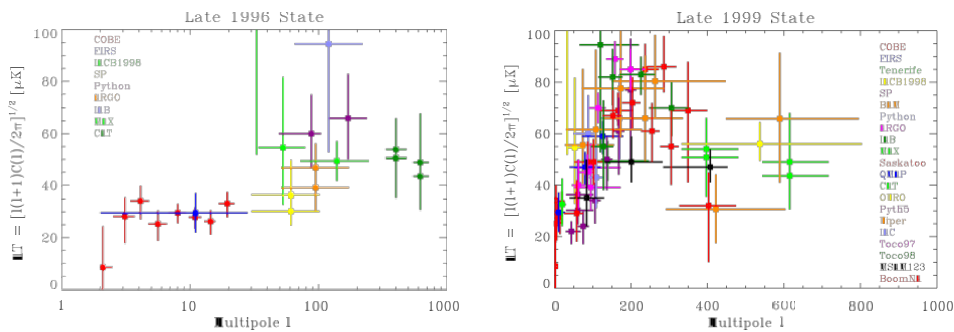

**Fig. 0.2** Successive early measurements of the temperature anisotropies angular power spectrum. a) The left panel shows all published detection at the end of 1996, the year when *Planck* was selected by ESA, while b) the plot at right is an update at the end of 1999.

### 0.1.1 The *Planck* challenge

Given the *COBE*-DMR results and the existing theoretical framework, one could then imagine (back in 1992) the experiment which we would like to do, based on physics, and that gave rise to *Planck*. Indeed, the measurement goals of *Planck* may be stated rather simply: *to build an experiment able to perform the "ultimate" measurement of the primary CMB temperature anisotropies*. This entails:

- a coverage of the entire sky and a good enough angular resolution in order to mine all scales at which the Cosmic Microwave background (CMB) primary anisotropies contain information;
- a very large frequency coverage, to allow removing precisely the astrophysical foreground contributions superimposed with variable strengths to the primary signal emission at any single measurement frequency.

We translated these into the high-level requirements for *Planck* to map the entire sky at 9 frequencies from 30 GHz to 1 THz, with an angular resolution and sensitivity at each of the survey frequencies in line with the role of each map in determining the CMB properties, in order to reach in the end of $\sim$ μK CMB sensitivity and an angular resolution of $\gtrsim 5$ minutes of arc[2]. For the measurement of the polarisation of the CMB anisotropies, *Planck* goal was "only" to get the best polarisation performances with the technology available at the design time[3].

Table 0.1 summarises the main performance goals of *Planck*, both in angular resolution and sensitivity expressed as the average detector noise, $c_{noise}$, within a square patch of 1 degree of linear size, for the 14 months baseline duration of the mission. This duration allows covering twice all the sky pixels (of size FWHM of Table 0.1) by nearly all the detectors. These goals are extracted from the 2004 *Planck* "Blue Book" issued to provide a complete overview of the planed scientific program (cf. url http://www.rssd.esa.int/SA/PLANCK/docs/Bluebook-ESA-SCI(2005)1_V2.pdf), 5 years before launch.

This is on these simple but ambitious goals (and on the proposed way of reaching them) that, after 3 years of preparatory work, the project was selected by the European Space Agency (ESA), as the $3^{rd}$ Medium size mission of its Horizon 2000+ program. This selection occurred in March 1996, *i.e.* contemporaneously with that of *WMAP* by NASA, which rather proposed reaching earlier less ambitious goals which could be attained with only incremental development on then existing technology.

In order to achieve the ambitious sensitivity goals of *Planck*, we proposed the HFI instrument, using a small number of detectors, limited principally by the photon noise of the background (for the CMB ones), in each frequency band. HFI stands simply for "High Frequency Instrument". This concept implied to achieve several technological feats never achieved in space before:

- sensitive & fast bolometers with a Noise Equivalent Power $< 2 \times 10^{-17}$ W/Hz$^{1/2}$ and time constants typically smaller than about 5 milliseconds (which thus requires to cool

---

[2]One can for instance show that the decrease with increasing angular resolution of the uncertainties of the cosmological parameters of the $\Lambda$CDM model levels off by $\ell \sim 1800$. Additionally, at substantially smaller scales, secondary fluctuations - if anything form foregrounds - completely dominate the primary ones.

[3]In the course of time, and with the successful developments of enabling technologies, we boosted our initial polarisation goals and set out to reach the polarisation sensitivity levels described in Table 0.1.

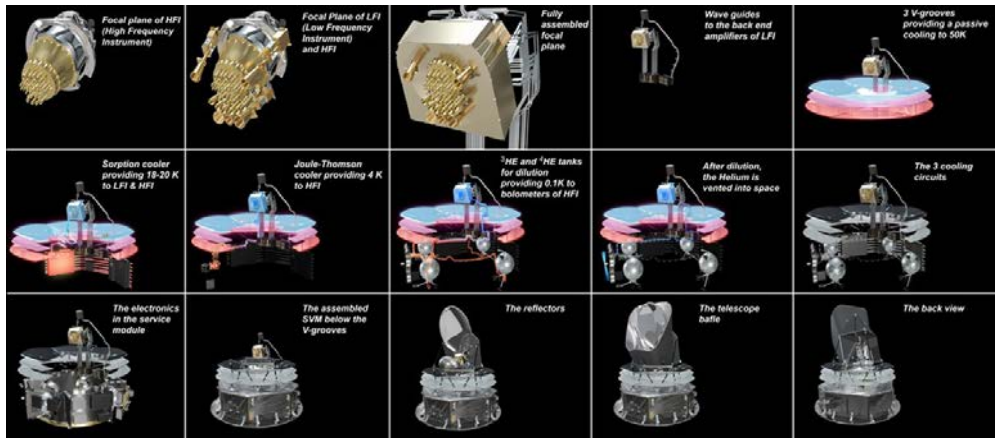

**Fig. 0.3** *Planck* build-up from the inside out. Going from left to right, one sees on the top row (t1) the HFI instrument with its 52 detector horns poking out of it's outer shell at 4 K. (t2) HFI surrounded by the 11 larger horns from the LFI. HFI and LFI together form (t3) the focal plane assembly from which (t4) the electrical signal departs (though a bunch of wave guides for the LFI and a harness of wires for HFI) to connect to the warm electronics parts of the detection chain which are located within the service module at ~ 300 K. (t5) The cold (top) and warm (bottom) parts are separated by three thermally isolating V-grooves which allow radiating to space heat from the spacecraft sideways and quite efficiently. The third (top) V-grooves operating temperature is about 40 K.

On the middle row, one sees (m1) the beds of the "20 K" sorption cooler and it's piping around the V-groove, bringing the overall focal-plane structure to LFI's operational temperature of ~ 18 K. (m2) The back-to-back (to damp the first harmonics of the vibrations) compressors of the 4 K cooler allow bringing HFI outer shell to 4 K, while (m3) isotopes from the $He^3$ tank and the three $He^4$ tanks are brought to the mixing pipes within HFI to cool filters (within the horns) to 1.6 K and the bolometer plate to 0.1 K, before (m4) being released to space. (m5) The passive cooling and the three active stage constitute this complex but powerful cooling chain in space.

On the bottom row, one can also see (b1) some of the electronic boxes in the service module (SVM) which in addition to the warm part of the electronic and cooling chains also contain all "services" needed for transmitting data, reconstructing the spacecraft attitude, powering the whole satellite... The bottom of the SVM is covered with solar panels, while supporting struts begin on its top which allows positioning (b3) the secondary and primary reflectors. The top part is surrounded by a large baffle to shield at best the focal plane from stray-light (b4). The back view (b5) allows distinguishing in the back the supporting structure of the primary mirror, and the wave guides from LFI. The spin axis of *Planck* (vertical on these plots) is meant to always remain close to the sun-earth line, with the solar panel near perpendicular to that line and the rotation of the line-of-sight (at 1 rpm) causing the detectors to survey circles on the sky with an opening angle around 85 degrees. Copyright ESA.

**Table 0.1** Summary of *Planck* performance. The *goals* we set were for the required 14 months of routine operations, with *requirements* on sensitivity two times worse than the stated goals. The (sky-averaged) sensitivities, $c_{noise}^{X}$, with $X = T$, $Q$ or $U$ indicate the *rms* detector noise, expressed as an equivalent temperature fluctuation in µK, which is expected once it is averaged in a pixel of one degree of linear size (*e.g.,* multiply by FWHM/60 to get the *rms* of the noise per pixels of size FWHM, if the noise is approximately white). The first 3 frequencies corresponds to the goals of the low frequency instrument, LFI, while the 6 other ones are covered by the high frequency instrument, HFI. For ease of comparison, the *actual* characteristics of the 2013 data release are also given here. The HFI outperforming is significant since the 2013 release is for about the same duration (14.5 months) than the duration we initially required, and corresponds to about half of the total data finally acquired by HFI.

| $\nu$ | [GHz] | 30 | 44 | 70 | 100 | 143 | 217 | 353 | 545 | 857 |
|---|---|---|---|---|---|---|---|---|---|---|
| | | \multicolumn{3}{LFI goals} | | \multicolumn{6}{HFI goals} | | | | | |
| FWHM | [arcmin] | 33 | 24 | 14 | 9.5 | 7.1 | 5.0 | 5.0 | 5.0 | 5.0 |
| $c_{noise}^{T}$ | [µK deg] | 3.0 | 3.0 | 3.0 | 1.1 | 0.7 | 1.1 | 3.3 | 33 | 1520 |
| $c_{noise}^{Q\,or\,U}$ | [µK deg] | 4.3 | 4.3 | 4.3 | 1.8 | 1.4 | 2.2 | 6.8 | | |
| | | \multicolumn{3}{LFI 2013} | | \multicolumn{6}{HFI 2013} | | | | | |
| FWHM | [arcmin] | 33.16 | 28.09 | 13.08 | 9.59 | 7.18 | 4.87 | 4.70 | 4.73 | 4.51 |
| $c_{noise}^{T}$ | [µK deg] | 5.1 | 5.9 | 5.1 | 1.8 | 0.7 | 1.0 | 3.4 | … | … |

them down to ≈ 100 mK, and to build them with a very low heat capacity & sensitivity to charged particles - more on this latter);
- total power read out electronics with very low noise, < 6 nV/Hz$^{1/2}$ in a large frequency range, from 10 mHz (1 rpm) to 100 Hz (*i.e.* from the largest to the smallest angular scales to measure at the *Planck* scanning speed);
- excellent temperature stability, from 10 mHz to 100 Hz, such that the induced variation would be a small fraction of the detector temporal noise (cf. Lamarre *et al.* 2010 for details):
  * better than 10 µK/Hz$^{1/2}$ for the 4K box (assuming 30% emissivity);
  * better than 30 µK/Hz$^{1/2}$ on the 1.6K filter plate (assuming a 20% emissivity);
  * better than 20 nK/Hz$^{1/2}$ for the detector plate (a damping factor ~ 5000 needed).

The other proposed instrument, LFI (for Low Frequency Instrument), required very low noise HEMT amplifiers (therefore cooled to 20 K) and very stable cold reference loads, at 4 K. In addition to the prowesses needed by the two focal plane instruments, *Planck* also demanded:
- a low emissivity telescope with very low side lobes (*i.e.* strongly under-illuminated);
- no windows, and minimum warm surfaces between the detectors and the telescope;
- a quite complex cryogenic cooling chain, which is illustrated in Figure 0.3. This chain begins with reaching ~ 40 K via passive cooling, by radiating about 2 Watts to space, followed by three active stages, at about 20 K, 4 K, and 0.1 K:
  * 20 K for the LFI, with a large cooling power, ~ 0.7 Watts, provided by $H_2$ Joule-Thomson sorption pumps developed by JPL, USA;

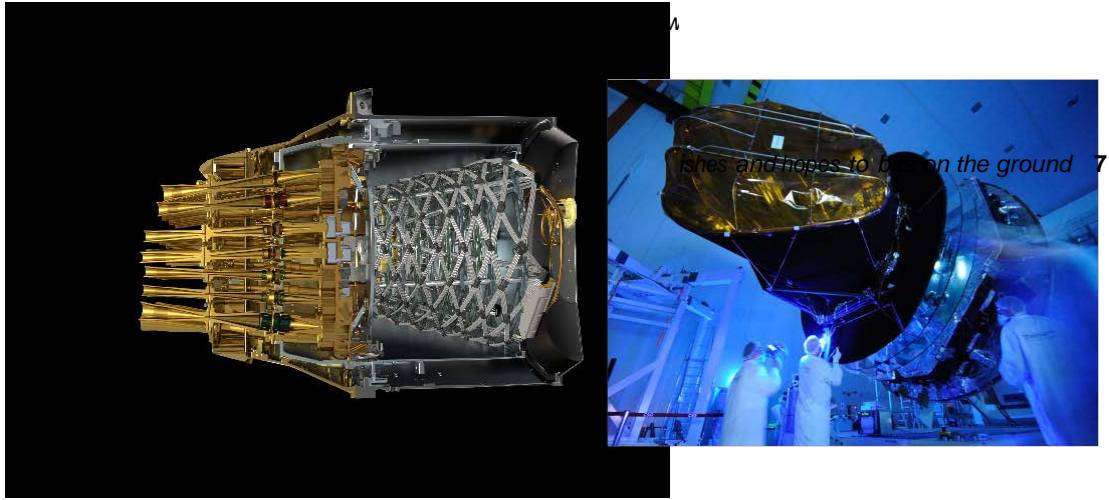

**Fig. 0.4** a) The cut-out at left displays the Russian doll arrangement of the HFI. Starting from the left, one sees the back-to-back horns going through HFI outer shell at 4 K, the filters fixed on the 1.6 K inner shell and shown in different colours according to their central frequency, followed by the horn on the top of the bolometers encasing at 0.1 K. These encasings are fixed on a plate at the end of the "basket weaving" of the electrical wires and of the dilution piping. It is within these pipes that the mixing of the He$^3$ and He$^4$ takes place and lowers the temperature from the 1.6 K shell at the right end to the 0.1 K of the bolometer plate at the other end. b) The picture of *Planck* at right was taken in 2009 in Kourou when it was time to "dust if off" as part of the preparations for launch.

* 4 K, 1.6 K and 100 mK for the HFI: the 15 milli-Watts cooling power at 4 K is supplied by mechanical pumps provided by the RAL, UK, in order to perform a Joule-Thomson expansion of He; the 1.6 K stage has a pre-cooling power of about 0.5 milli-Watts, thanks to another Joule-Thomson expansion, while the final dilution fridge of He$^3$ & He$^4$, from a French collaboration between Air Liquide and the CRTBT, can lift 0.2 micro-Watts at 0.1 K;
* a thermal architecture optimised to damp thermal fluctuations (active+passive).

Furthermore, a tight control of vibrations is needed, in particular since the dilution cooler does not tolerate micro-vibrations at sub-mg level. And as little as $7 \times 10^{10}$ He atoms accumulated on the dilution heat exchanger (an He pressure typically at the $1 \times 10^{-10}$ mb level) would be too much. Fig. 0.4-a shows how HFI was designed to reach these objectives.

These top-level design goals were turned into real instruments, which went through several qualification models. The HFI instrument (Lamarre *et al.*, 2010; Planck HFI Core Team, 2011*a*) comprises 52 signal bolometers, as well as two dark bolometers, 16 thermometers, a resistor, and a capacitor used for monitoring and housekeeping[4]. The bolometers includes twelve polarization sensitive bolometer (PSB) pairs, four each at 100–353 GHz; the rest are unpolarized spider-web bolometers (SWBs). The LFI instrument (Bersanelli *et al.*, 2010; Mennella *et al.*, 2011) detectors are 22 pseudo-correlation radiometers, covering three bands centred at 30, 44, and 70 GHz[5].

Before delivering the actual flight model of both instruments to industry for integration with the satellite, both instrumental consortia organised extensive calibrations campaigns, starting at the individual components levels, then at the sub-systems levels (*e.g.,* individual photometric pixels), then at instrument level. For HFI, the detector-level tests were done mainly at JPL in the USA, and the pixel level tests were performed in Cardiff in the UK, while the instrument calibration was performed at the Institut d'Astrophysique Spatiale in Orsay, France from April till the end of July 2006. During that period, we obtained in particular 19 days of scientific data at normal operating conditions. We could then confirm that HFI satisfied all our requirements, and for the most part actually reached or exceeded the more ambitious design goals, in particular concerning the sensitivity, and speed of the bolometers, the very low noise of the read out electronics and the overall thermal stability. The LFI instrument also went through detailed testing around the same time and it reached many of its ambitious requirements.

---

[4]We actually include in the analysis the signal from 50 bolometers only, since two spider-web detectors are currently not used due to their high cosmic ray sensitivity.

[5]The initial design of LFI included a large number of detectors at 100 GHz, the inclusion of which was unfortunately abandoned for programmatic and financial reasons.

The integration of the LFI and HFI instruments was performed at the premises of the prime industrial contractor, Thales, in Cannes in November 2006. Within a year, by December 2007, the full satellite was ready for vibration testing. *Planck* was then flown from Cannes to ESA's ESTEC centre (in Noordwijk, Holland) where among other things it went through load balancing on April $7^{th}$ 2008, before travelling again to the "Centre Spatial de Lièges" (CSL). This ultimate system-level (ground) test, with all elements of the cryogenic chain present and operating, demonstrated in particular the following: a) the dilution system can work with the minimal Helium 3 and 4 flux, which let us hope 30 months of survey duration (as it actually happened, nominal duration being 14 months!). b) the extremely demanding temperature stability required (at 1/5 of the detection noise) has been verified, c) bolometers sensitivities in flight conditions are indeed centred around their goal.

*Planck* was then shipped to Kourou, prepared for launch (see fig. 0.4-b), and after a few more nerve-racking delays, we finally lost sight of *Planck* for ever (when it was covered by the SYLDA support system on the top of which laid *Herschel* for a joint launch). Launch was on May $14^{th}$ 2009, and it was essentially perfect. After separating from *Herschel*, *Planck* was set in rotation and started its travel to the L2 Lagrange point of the sun-earth system. L2 is at 1.5 million kilometres away from earth, *i.e.*, about 1% further away from the sun than the earth. The final injection in the L2 orbit was performed at the end of June 2009 (see figure 0.5-a), at the same time than the cooling sequence ended successfully. Indeed, figure 0.5-b shows how the various thermal stages reached their operating temperature, cooling of course from the outside-in, and closely following the predicted pattern.

Once at L2, a calibration and performance verification phase was conducted till mid-august, to insure that all system were working properly and that instrumental parameters were all set at best. From August $13^{th}$ to $27^{th}$, we conducted a "First Light Survey" in normal operational mode for an ultimate verification of the parameters and of the long-term stability of the experiment. We found the data quality to be excellent, and the Data processing Centre pipelines were successfully operated as hoped. Indeed the first maps were produced within days of getting the data (see Fig. 0.6-a), and clearly showed consistency of the mapping of the CMB component by the two instruments.

Since then, the operations have been extremely smooth, the instruments have been *very* stable, as can be judged from the temperature temporal records displayed in Fig. 0.6-b. Note in particular that the 0.1 K stage of the bolometers plate has been stable at better than a part in a thousand over the duration of the first 4 surveys. The few spikes at the very end can be traced to specific spacecraft events. Bolometers were thus regulated at the 0.00001 K level over

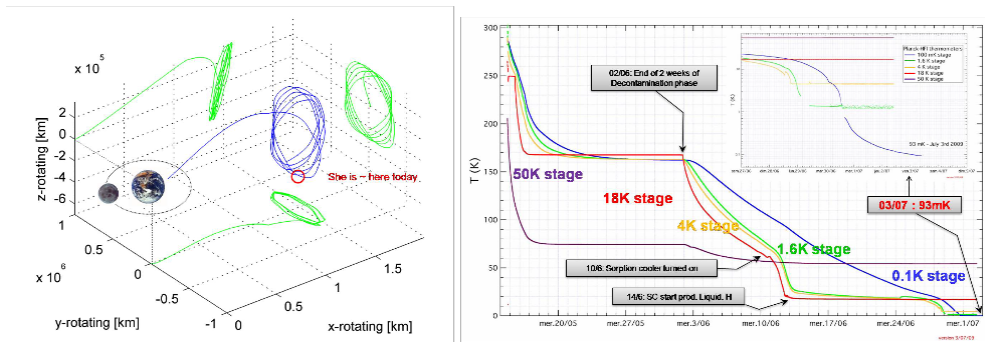

**Fig. 0.5** a) Spacecraft trajectory to and on the L2 orbit. b) Cooling sequence of *Planck*, showing the various stages reaching in turn their operational temperature, till the dilution plate actually reached 93 mK on July 3$^{rd}$. Credit ESA and HFI consortium.

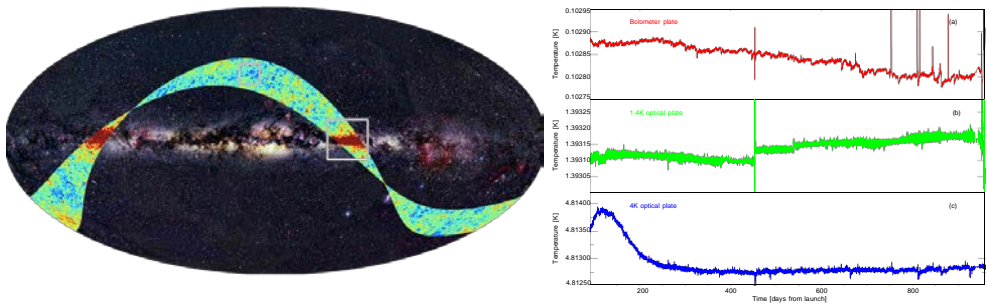

**Fig. 0.6** a) The first light survey map, obtained on the basic of the first 2 weeks of survey data gathered in August, was made public in september 2009. It demonstrated that *Planck* was "fit for service". b) Temperature variations of critical stages over the full mission duration. Credit ESA & HFI consortium.

nearly 900 days, a feat which played a major role in the final quality of HFI sky maps! This is not to say though that everything went according to plan, nor that absolutely all had been planned beforehand. Indeed, for HFI, the data processing teams had to deal with a number of in-flight surprise to which we shall return later (the abundance of cosmic ray hits, the ADC non-linearity, the contribution of CO lines, etc.). Still, the instrumental teams and their associated Data Processing Centre (DPC hereafter) have successfully improved on their planned data processing to create the main legacy of the mission, nine sky maps of unprecedented combination of coverage, sensitivity, resolution, and accuracy which are shown in Fig. 0.7. They constitute the primary legacy of the mission.

## 0.2 From bits to maps

I now describe steps taken to transform the packets sent by the satellite into sky maps with the help of ancillary data, for example, from ground calibration. The 2013 sky maps are intensity maps alone, as obtained from the beginning of the first light survey on 13 August 2009, to the end of the nominal mission on 27 November 2010. The overview given below corresponds to the data handling for this release by the HFI Data Processing Centre (hereafter DPC) whose development and operation has been a major involvement of mine over more than 10 years (Planck HFI Core Team, 2011*b*; Planck Collaboration VI, 2013). While specific to HFI, it illustrates well the complexity of the process which led to apparently simple sky maps, and hopefully makes clear why despite this complexity the cosmology results based on the resulting data appear to be sound and robust.

The processing of HFI data proceeds according to a series of levels, shown schematically in Fig. 0.8. Level 1 (L1) creates a database of the raw satellite data as a function of time (TOI objects, for time-ordered information). The full set of TOIs comprises the signals from each HFI bolometer, ancillary information (e.g., pointing data), and associated housekeeping data (e.g., temperature monitors). Level 2 (L2), the main subject of this section, uses these data to build a model of the HFI instrument, the so-called "Instrument Model" or IMO, produces cleaned, calibrated time-lines for each detector, and combines these into aggregate products such as maps at each frequency and their characteristics. Level 3 (L3) takes these instrument-specific products and derives further refined scientific products: component-separation algorithms transform the maps at each frequency into maps of separate astrophysical components; source detection algorithms create catalogues of Galactic and extragalactic objects; finally, a likelihood code assess the match between a cosmological and astrophysical model and the frequency maps. Some of the L3 processing is described in the following Sect. 0.3.

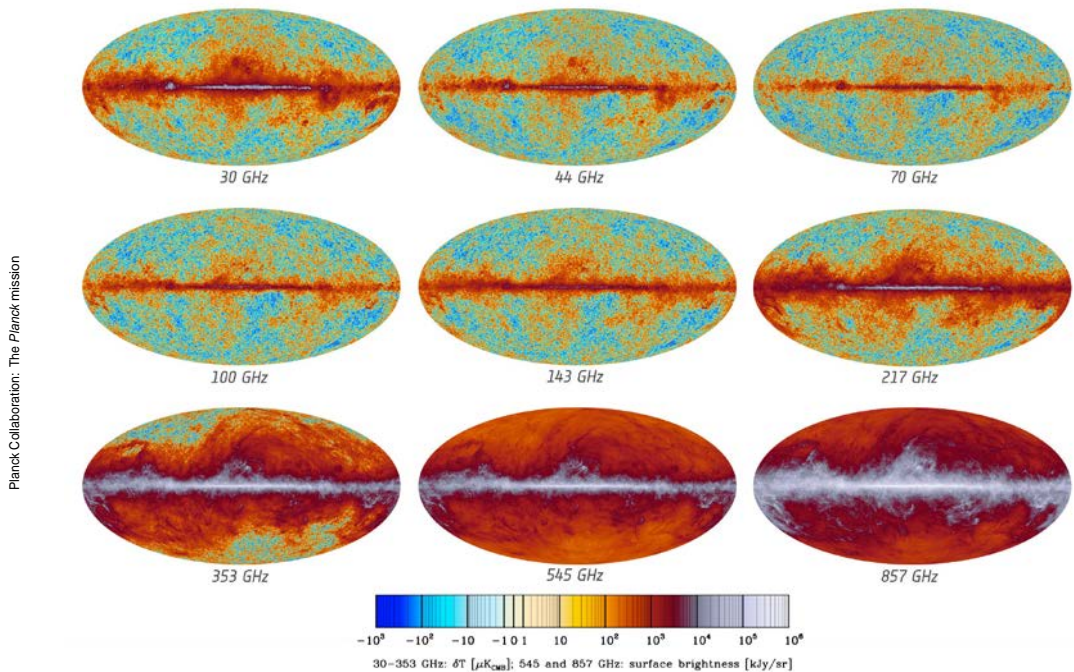

**Fig. 0.7.** The nine Planck frequency maps show the broad frequency response of the individual channels. The color scale based on inversion of the function $y = 10^x - 10^{-x}$, is tailored to the full dynamic range of maps from *Planck* nominal mission data. Credit ESA and HFI & LFI consortia.

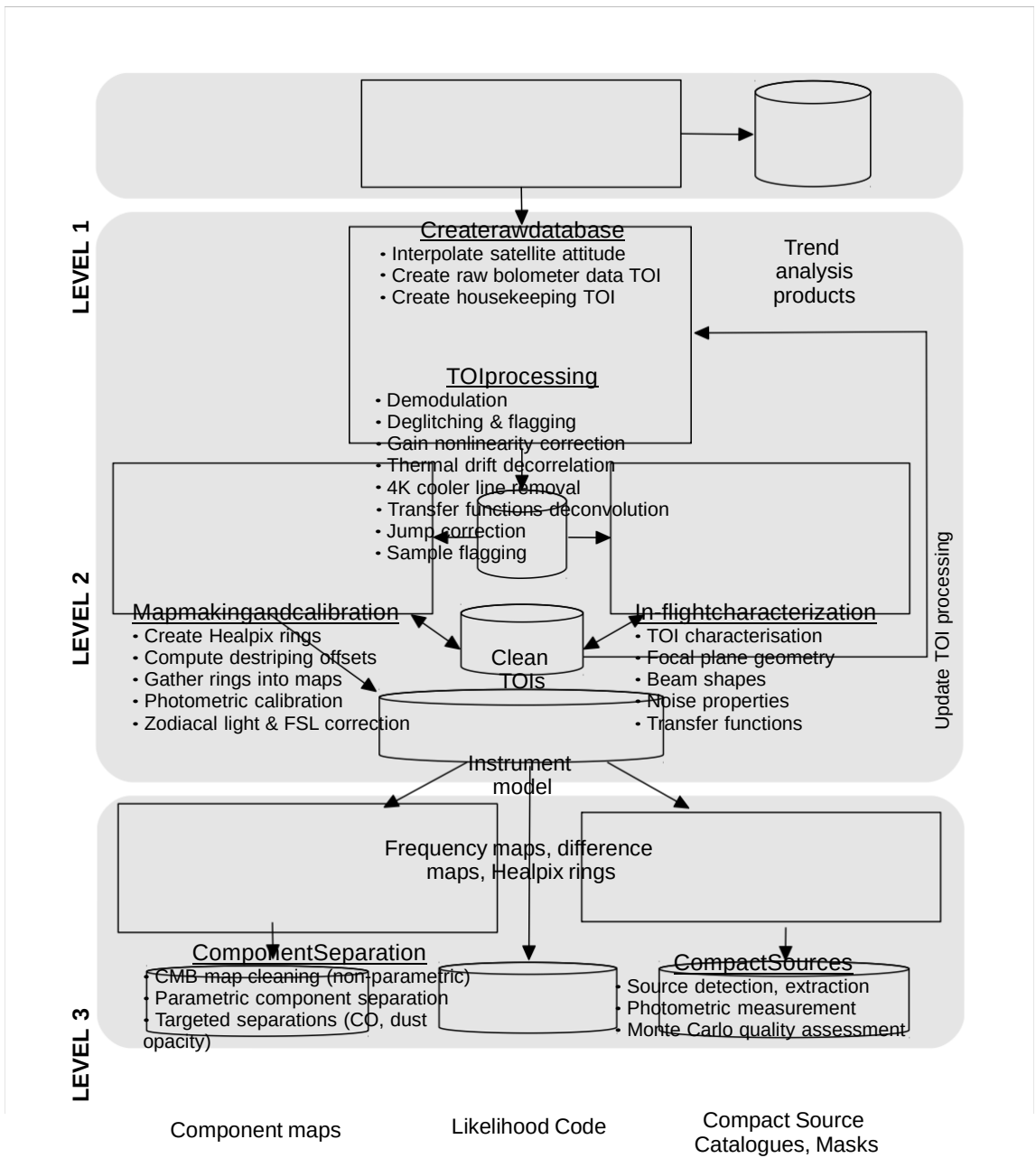

**Fig. 0.8** Overview of the data flow and main functional tasks of the HFI Data Processing Centre. Level 1 creates a database of the raw satellite data as a function of time. Level 2 builds a data model and produces maps sky maps at the six frequency of the instrument. Note the crucial role of the Instrument Model, which is both an input and an output of many tasks, and is updated iteratively during successive passes of the data. Level 3 takes these instrument-specific products and derives final astrophysical products.

Of course, these processing steps are not done completely sequentially: HFI data are processed iteratively. In many ways, the IMO is the main internal data product from *Planck*, and the main task of the HFI DPC is its iterative updating. Early versions of the IMO were derived from pre-launch data, then from the first light survey of the last two weeks of August 2009. Further revisions of the IMO, and of the pipelines themselves, were then derived after the completion of successive passes through the data. These new versions included expanded information about the HFI instrument: for example, the initial IMO contained only coarse information about the shape of the detector angular response (i.e., the full-width at half-maximum of an approximate Gaussian); subsequent revisions also included full measured harmonic-space window functions.

In somewhat more detail, L1 software fills the database and updates (daily during operations, while they lasted), the various TOI objects. Satellite attitude data, sampled at 8 Hz during science data acquisition and at 4 Hz otherwise, are re-sampled by interpolation to the 180.37370 Hz (hereafter 180.4 Hz) acquisition frequency of the detectors, corresponding to the integration time for a single data sample; further information on L1 steps were given in (Planck HFI Core Team, 2011*b*).

Raw time-lines and housekeeping data are then processed by L2 to compensate for instrumental response and to remove estimates of known artefacts. First, the raw time-line voltages are demodulated, deglitched, corrected for the bolometer non-linearity, and for temperature fluctuations of the environment using correlations with the signal TOI from the two dark bolometers designed as bolometer plate temperature monitors. Narrow lines caused by the $^4$He-JT (4 K) cooler are also removed before deconvolving the temporal response of the instrument. Finally, various flags are set to mark unusable samples. Fig. 0.9 illustrates the effect of this TOI processing. The most striking effect is the removal of a large number of "glitches", resulting from the impact of cosmic rays on the detectors and their encasing. These hits create showers of secondaries which induce many "glitches" in the data flow of some bolometers and also relatively slow temperature fluctuations of the bolometer plate which create an additional noise correlated among all detectors. The variability and number density of these glitches had not been fully anticipated before launch, and they constituted one of the "surprises" we had to mitigate in the data processing. Fig. 0.10 illustrates visually our glitch removal and flagging process, by zooming in a short section of temporal data.

Further use of the data requires knowledge of the pointing for individual detectors. During a single stable pointing period, *Planck* spins around an axis pointing towards a fixed direction on the sky (up to an accounted-for wobbling), repeatedly scanning approximately the same circle (Planck Collaboration I, 2013). The satellite is re-pointed so that the spin axis follows the Sun, and the observed circle sweeps through the sky at a rate of approximately one degree per day. Assuming a focal plane geometry, i.e., a set of very slowly varying relations between the satellite pointing and that of each of the detectors, we build rings of data derived by analysing the data acquired by a detector during each stable pointing period ("ring" refers to the data obtained during a single stable pointing period). This redundancy permits averaging of the data on rings to reduce instrument noise. The resulting estimate of the sky signal can then be subtracted from the time-line to estimate the temporal noise power spectral density, a useful characterization of the detector data after TOI processing. This noise may be described as a white noise component, dominating at intermediate temporal frequencies, plus additional low- and high-frequency noise. The effect on maps of the low frequency part of the noise can be partially mitigated by determining an offset for each ring. These so-called "destriping" offsets are obtained by requiring that the difference between intersecting rings be minimized. Once the offsets are removed from each ring, the rings are co-added to produce sky maps[6].

---

[6] A correction of the rings for Zodiacal light emission was implemented, but in the end it was found that component separation method did a better job than our correction at removing that emission from the CMB map.

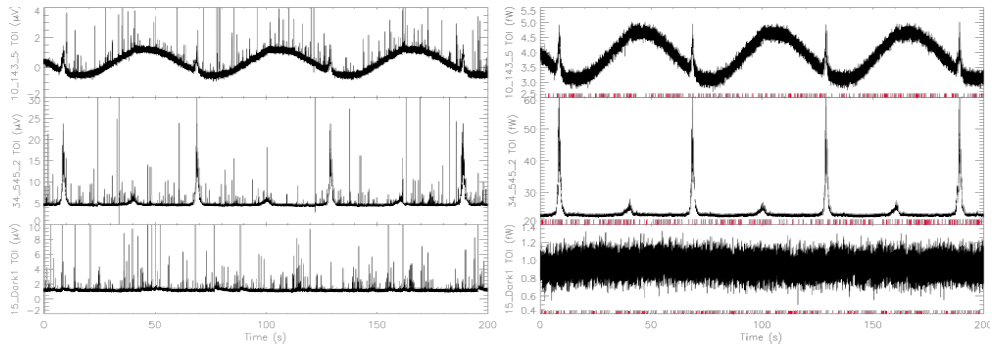

**Fig. 0.9** a) Raw TOIs for three bolometers, the "143-5" (top), "545-2" (middle), and "Dark1" (bottom) illustrating the typical behaviour of a detector at 143 GHz, at 545 GHz, and of a blind detector over the course of three rotations at 1 r.p.m. of the spacecraft. At 143 GHz, one clearly sees the CMB dipole with a 60 s period. The 143 and 545 GHz bolometers show vividly the two Galactic Plane crossings, also with 60 s periodicity. The dark bolometer exhibits a nearly constant baseline together with a population of glitches from cosmic rays similar to those seen in the two upper panels. b) Processed TOI for the same bolometers and time range as shown in a). Times where data are flagged (unusable), are indicated by the purple ticks at the bottom of each plot. Credit HFI consortium, (Planck HFI Core Team, 2011*b*).

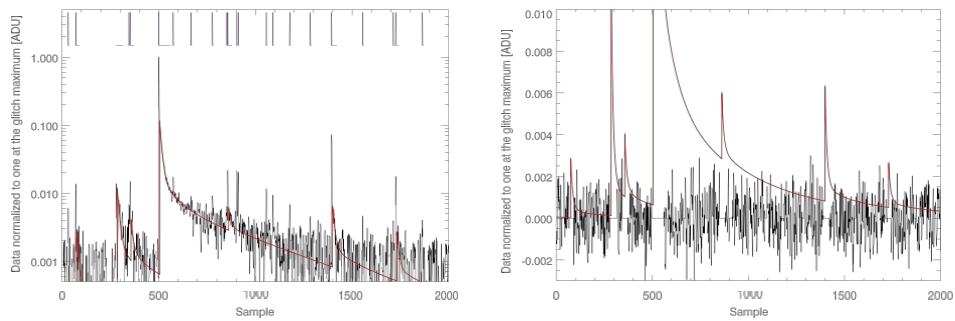

**Fig. 0.10** a) Example of 2000 samples of sky subtracted data encompassing a large event in black and our best fit glitch templates (red). The purple ticks in the upper part of the figure show where data are flagged and indicate the detected position of glitches. b) The cleaned residual in black (with the flagged areas set to zero) compared with the fitted template in red.

A complication arises from the fact that the detector data include both the contribution from the Solar dipole induced by the motion of the Solar System through the CMB (sometimes referred to as the "cosmological" dipole), and the orbital dipole induced by the motion of the satellite within the Solar System, which is not constant on the sky and must therefore be removed from the rings before creating the sky map. The Solar dipole is used as a calibration source at lower HFI frequencies[7], and bright planet fluxes at higher frequencies. Since we need this calibration to remove the orbital dipole contribution to create the maps themselves, the maps and their calibrations are obtained iteratively. The dipoles are computed in the non-relativistic approximation. The resulting calibration coefficients are also stored in the IMO, which can then be used, for instance, to express noise spectra in noise equivalent temperature (NET).

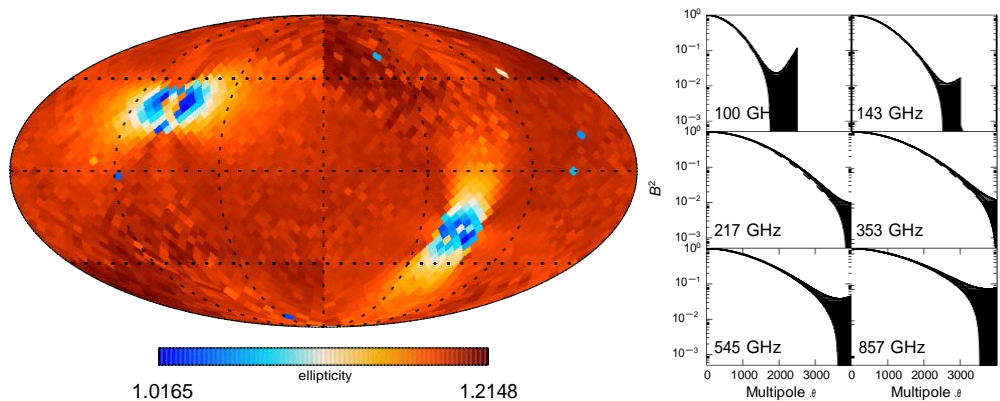

**Fig. 0.11** Beam properties. a) The best-fit Gaussian ellipticity of the 100 GHz effective beam across the sky in Galactic coordinates. The beams are clearly isotropised in the high redundancy regions close to the ecliptic poles, when scans with many directions are used jointly to make the map. b) Effective beam window functions (solid lines) for each HFI frequency. The shaded region shows the full ±1σ error envelope. Dashed lines show the effective beam window function for Gaussian beams with full-width at half-maximum (FWHM) parameter $9.\!^t65$, $7.\!^t25$, $4.\!^t99$, $4.\!^t82$, $4.\!^t68$, and $4.\!^t32$ for 100, 143, 217, 353, 545, and 857 GHz respectively).

The destriping offsets, once obtained through a global solution, are also used to create local maps around planets. These are used to improve our knowledge of the focal plane geometry stored in the previous version of the IMO and to improve measurements of the "scanning" beam", which is defined as the beam measured from the response to a point source of the full optical and electronic system, after the filtering done during the TOI processing step. This response pattern therefore includes, in addition to the optical response, any further effect due to the imperfections of the temporal data processing. Let us note though that the map-making procedure implies that any map pixel is the sum of many different samples in the time-line, each of which contributing to the pixel in a different location and with a different scan direction. Thus, we further define the "effective" beam of each map pixel which takes into account the details of the scan pattern. An example of positional variation of the effective beam (it's ellipticity at 100 GHz) is given in Fig. 0.11-a). Finally, the multiplicative effect on the angular power spectrum is encoded in the effective beam window function (Hivon *et al.*, 2002), which includes the appropriate weights of each multipole for analysing aggregate maps (possibly masked) across detector sets or frequencies. Fig. 0.11-b) displays the effective beams of each HFI frequency channel and the associated uncertainties, as well as the best fitting Gaussian for that data combination.

---

[7] The pre-flight plan had been for HFI to use the perfectly well-known albeit weaker, orbital dipole, as an absolute calibrator, and deduce an improved solar dipole rather than calibrate on it. But we found that residual systematics in the processed TOIs lead to instabilities preventing us to follow that route for now.

The ring and map-making stages allow us to generate many different maps, e.g., using different sets of detectors, the first or second halves of the data in each and every ring, or from different sky surveys (i.e., acquired during different periods). Null tests using difference maps have proved extremely useful in revealing systematic effects and characterizing the maps. Figure 0.12 shows on top the 143 GHz map, which can be compared with the difference between maps made of the first and the second half of each stable pointing period (i.e., half-ring maps, middle), and the difference between survey-1 and survey-2 maps (bottom). These surveys were obtained during two consecutive 6 months periods of observation. The top plot has a scale restricted to 500 µK., therefore completely saturating the Galactic plane emission; still, even with a 100 times enlarged scale of 5 µK, the middle plot only shows what appears to be a noise pattern modulated by the observing time. While spectacular, this test map only displays the distribution of the collected signal which varies on a time scale shorter than about 20 minutes, by construction (a stable pointing period is typically 46 minutes long). The bottom plot shows survey differences, and is therefore sensitive to any variation on a time-scales shorter than 6 months. This very inclusive plot mainly reveals a residual within the Galactic plane, due to the fact that the strong Galactic plane signal is scanned six months apart with opposite directions, thereby revealing scanning beam asymmetries. But one also discerns further large scale modulation which needs a sharper tool to characterise.

In order to quantitatively assess these residuals, the *Planck* team uses Monte-Carlo simulations of the signal and noise. For all simulations, the sky signal is taken from the *Planck* Sky Model[8] (PSM), which has been described in detail in Delabrouille *et al.* 2012. The simulation pipeline then generates time-ordered data for the actual pointing of all the detectors by integrating the simulated sky signal within the frequency band-passes of each detectors, performing the convolution by their respective scanning beam response, as determined on the data too, and adding realistic noise realisations. This generates TOIs which, once flagged, are similar to the actual ones produced by the TOI processing step, if that stage has successfully removed all further instrumental effects[9]. These TOIs are then passed to the same map-making stage than the actual data, allowing to assess whether the null tests are acceptable given the data model we use for analysing them.

Figure 0.13 shows binned spectra $\ell(\ell + 1) C_\ell$ of difference maps made from various data combination. We actually use cross-spectra between independent detector sets (rather than auto-spectra) to allow for a direct comparison to the spectra used as input for the likelihood analysis which we shall described later (and which also avoids requiring a very precise simulation of the noise bias of auto-spectra). The residual signal in each band-power is shown up to the highest multipole used in the likelihood code (Planck Collaboration XV, 2013). For clarity, the bins used in the likelihood analysis are grouped four-by-four for $\ell > 60$ and eight-by-eight for $\ell > 1250$ in the plots. The Monte-Carlo simulations are used to remove from the measurement the mean of the simulations (which is very small due to using cross-spectra) and to affect as error bars the variance of the simulated $C_\ell$ distribution in each bin. The detector sets ds1 and ds2 are the two groupings of four polarised detectors (PSBs) available at 100, 143, and 217 GHz which we used to create (T, Q, U) maps per survey. We then created their survey-difference maps (Survey 2 − Survey 1)/2.

---

[8] This parametric model allows the generation of all-sky temperature and polarization maps of the CMB, the SZ effects, and diffuse Galactic emission (in particular synchrotron, free-free, and thermal dust) with a resolution of a few arc minutes at all *Planck* frequencies. The PSM also includes an extensive point-source catalogue, as well as spinning dust, CO line, and $H_{II}$ region models.

[9] Additional effects can of course be simulated and processed to verify that there are indeed negligible at the level of accuracy considered.

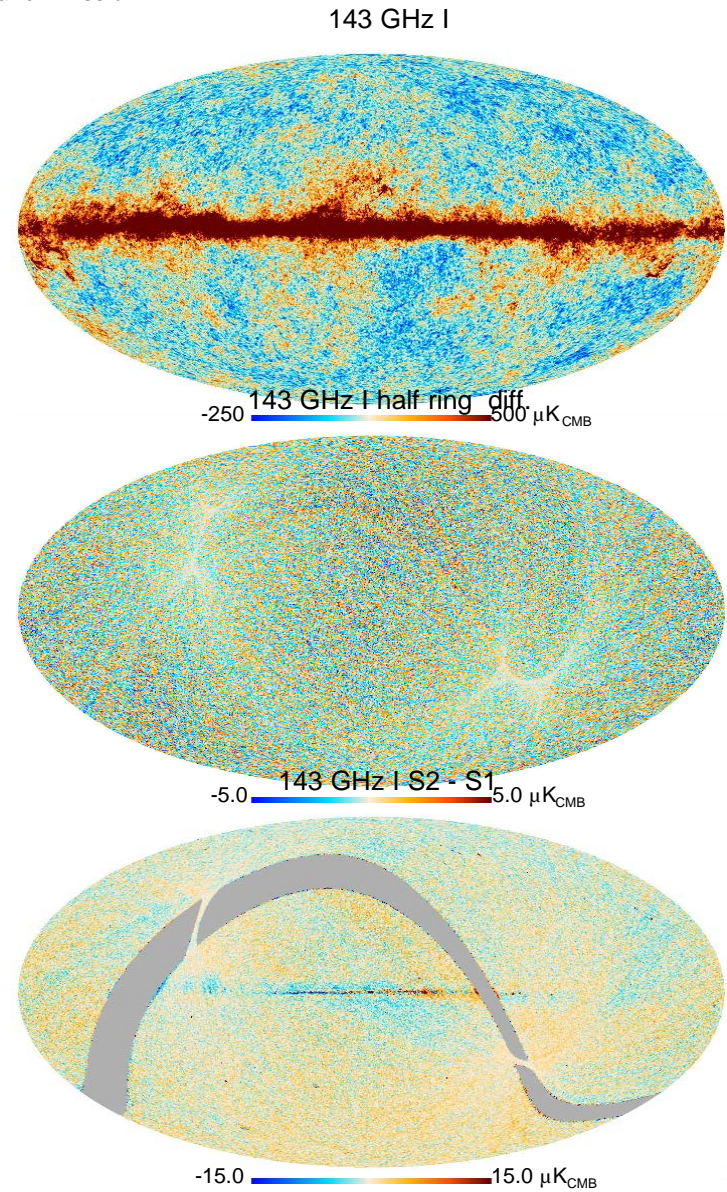

**Fig. 0.12** HFI maps at 143 GHz. The first row gives the intensity in $\mu K_{CMB}$. The second row shows the difference between maps made of the first and the second half of each stable pointing period (i.e., half-ring maps). The third row shows the difference between the first and second survey maps. The grey pixels were not observed in both surveys.

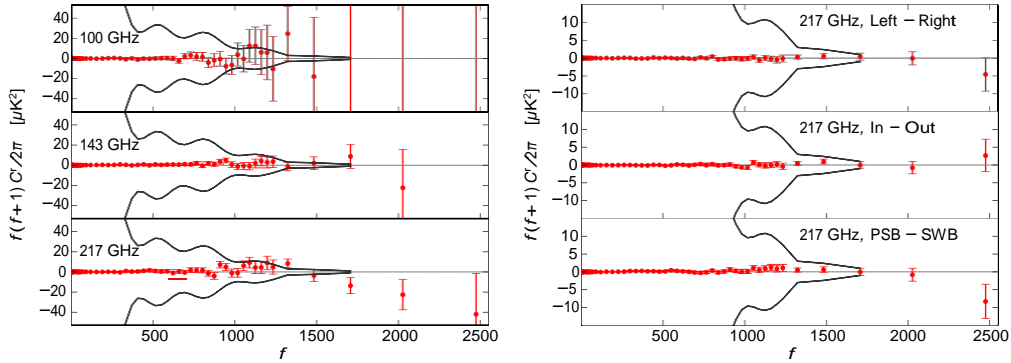

**Fig. 0.13** Consistency test result from some detector sets cross spectra, for the $f_{\rm sky} \simeq 0.30$ mask derived from that used in the primary cosmological analysis. We show in each bin the variance of the $C_\ell$ distribution from Monte-Carlo simulations. Note that the amplitude of the residual signal is (well) under the binned sample variance envelope expected for the $f_{\rm sky} \simeq 0.30$ mask (shown in black) up to $\ell \sim 1000$ in all cases. *Left*: the ds1 × ds2 cross spectra of (Survey 2 − Survey 1)/2 difference maps at 100, 143, and 217 GHz. *Right*: Other detector combinations at 217 GHz, grouping detectors on either side of the focal plane (top), or in concentric area (middle), or by polarisation capability (bottom).

Figure 0.13-a) shows the cross-spectra of the difference maps made from each of the two detector sets. At all three CMB frequencies, the $\ell < 200$ residuals are strikingly small, actually never exceeding about 0.5 $\mu K^2$. Although these non-zero differences are detected with very high statistical significance (confirming the visual impression of Fig. 0.12), they are many orders of magnitude smaller than the binned sample variance (shown in black in Fig. 0.13) at these scales and therefore irrelevant for the cosmological analysis of the temperature anisotropies. This stays true all the way up to $\ell \sim 1000$, at which point residuals become higher than the binned sample variance at 100 and 217 GHz. At 143 GHz, this does not happen until $\ell$ reaches 1500. In the multipole range from 1000 to 2500, although the amplitude of the residuals is almost always greater than the binned sample variance, the variance in the simulation results is significantly larger than in the $\ell < 1000$ regime. As a result, the 100 and 143 GHz residuals for that range of multipoles are fully compatible with zero. However, the 217 GHz residuals in the same multipole range are not, as can be directly inferred from the plot, where an oscillatory feature starts at $\ell \sim 1000$. But this is not seen in Fig. 0.13-b) for other data combinations at 217 GHz which pass well this nulling test. In any case, such difference tests have been performed for all possible combinations of the input maps used in the likelihood analysis. In addition to the 217-ds1 × 217-ds2 cross-spectrum, only two other cross-spectra fail this test; we have checked that the determination of the cosmological parameters is not

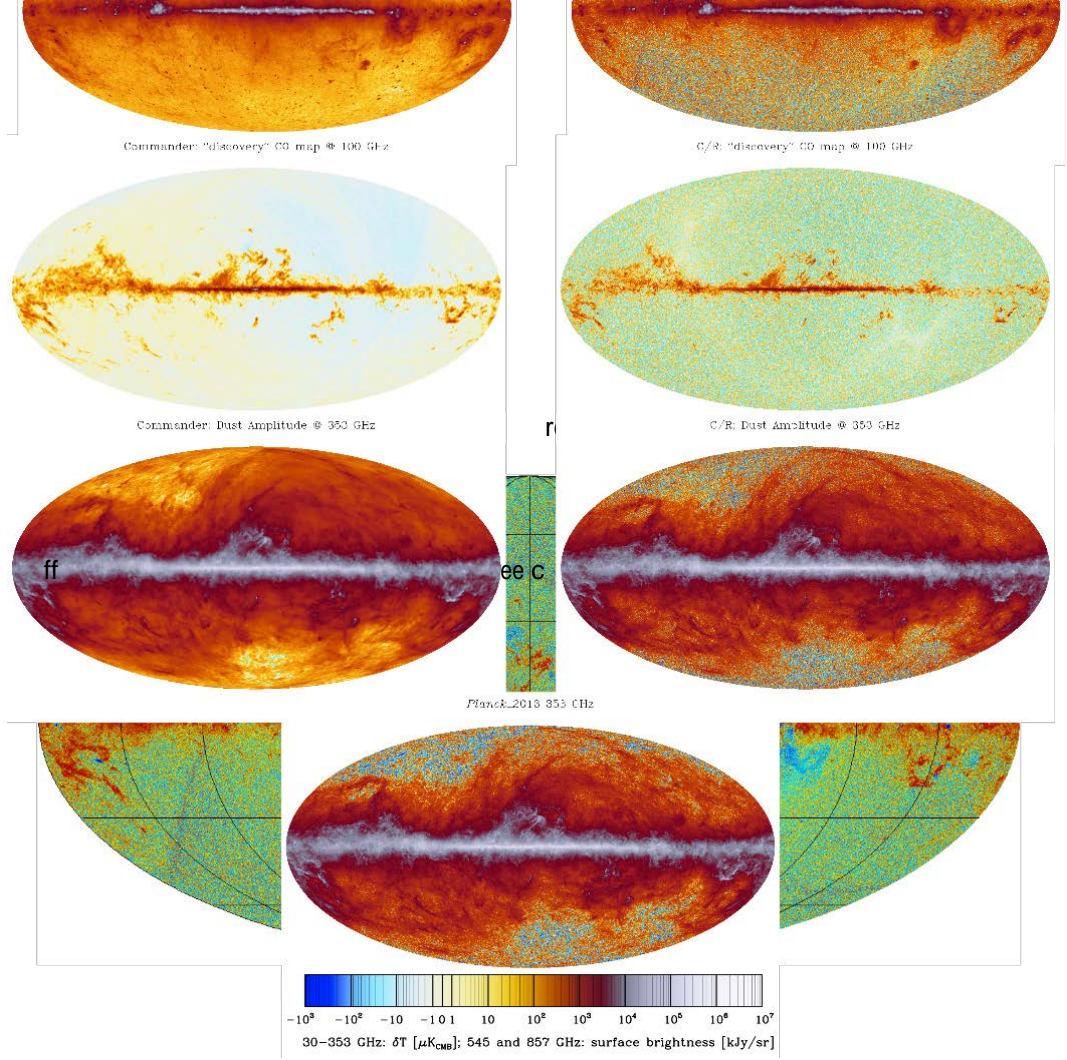

**Fig. 0.14** Difference of *Planck* 70 and 100 GHz maps expressed in equivalent thermodynamic temperature. This checks the nulling of CMB anisotropies determined by the two instruments. A good fraction of the Galactic emission which stands out at low latitudes arises from CO in the 100 GHz channel.

One of the key design features of *Planck* is that it contains two separate instruments, subject to independent calibration and systematic effects. The simple fact that they observe the same CMB anisotropies in nearly adjacent frequency bands, and that they do so with high signal-to-noise, provides a powerful cross-check on data quality. Figure 0.14 shows a map-level comparison between 70 and 100 GHz, the closest frequencies between the two instruments. The CMB structures at high Galactic latitude disappear in the difference made in thermodynamic temperature units as shown by the green (very close to zero) uniform background. The deep nulling of the CMB anisotropy signal directly achieved by this very straightforward differencing demonstrates that the Planck maps are free from serious large- to intermediate scale imperfections. It also reveals in an immediate and interesting manner the foreground residuals. However, the very high accuracy (~ 0.1%) which is the aim of *Planck* also implies that every minute difference in how the CMB anisotropies are observed must be taken into account when comparing data from LFI and HFI. This applies particularly to instrumental issues (beam shapes, noise levels) and residual foreground signals which are better assessed at the power spectrum level. This will be addressed in the next section.

While the *Planck* maps meet a very high standard, we did find limitations. Their mitigation, and related data products, are left to future releases. In particular, HFI analysis revealed that non-linear effects in the on-board analogue-to-digital converters (ADC) modified the recovered bolometer signal. *In situ* observations over 2012–2013 have measured this effect, and algorithms have been developed to explicitly account for it in the data analysis, albeit too late to include in the data processing for the 2013 release. However, it is important to understand that the first-order effect of the ADC non-linearity mimics a gain variation in the bolometers, which the 2013 release measures and removes as part of the calibration procedures. The HFI team has also identified another weak systematic effect (affecting the long term response of the bolometers) whose correction will likely allow an even higher calibration accuracy of the data and hopefully make possible a cosmological analysis of the large-scale polarisation (which did not pass the nulling test above and was therefore not part of the 2013 analyses).

## 0.3 From maps to CMB statistical characteristics

The nine all-sky high-sensitivity high angular resolution *Planck* maps are a treasure trove for astrophysics, which have already allowed many progresses in the understanding of the various astrophysical sources of emission in the millimetre and sub-millimetre range. Here I only focus on the CMB component, but let us still note that many exciting results have been obtained on the diffuse Galactic emission (in particular synchrotron, free-free, CO, spinning and thermal dust), as well as compact sources (radio-sources, Infra-red galaxies, Sunyaev-Zeldovich clusters) and the unresolved Cosmic Infra-red background. And these scientific exploitations just started, generating many follow-up studies with other facilities.

### 0.3.1 CMB map cleaning

In order to clean the background CMB map from foreground emissions, we have used four different approaches which combine differently the various frequency maps:

- a parametrised model approach in pixel space, Commander/Ruler, which was used to derive Galactic foregrounds maps, and the low-$\ell$ part of the likelihood code (for that, see § 0.3.2 next),
- a blind harmonic space approach, SMICA, which generated our reference map, in particular for CMB non-Gaussianity studies,
- a blind needlet space approach, NILC, which allows checking the benefits of spatial localisation,
- a spatial template based approach, SEVEM, which allows producing easily several CMB maps.

Different methods have different objectives and possibilities, in line with the specific problem they set out to solve best. Each component separation method produces at least a CMB map, a confidence map (i.e., a mask), an effective beam, and a noise estimate map characterising that CMB map.

Our first two algorithms are based on model fitting. Commander-Ruler (C-R) is a Bayesian parameter fitting approach which works in the pixel domain by fitting a parametrized model of the CMB and the foregrounds contribution to the data. The Commander part performs an MCMC sampling of the amplitudes and spectral parameters of the model at low resolution ($40^t$), while the Ruler part solves for high-resolution amplitudes using the Commander spectral parameters. The fit uses the maps of the 30 to 353 GHz channels. The end results have a $7\overset{t}{.}4$ resolution. Our other model fitting approach is SMICA, which performs spectral matching in the harmonic domain. It thus fits a model of the CMB, generalized correlated foregrounds, and noise contributions to the auto- and cross-spectra of all the maps from 30 to 857 GHz. The derived harmonic weights are then used to produce a CMB map with $5^t$ resolution. This map can be filled with constrained realization in the processing mask (3% of the sky), as in Fig. 0.15.

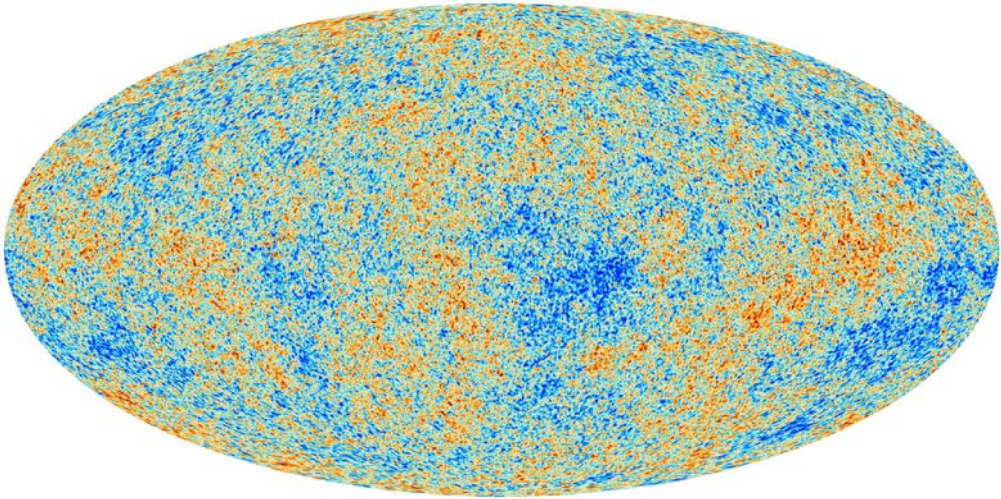

**Fig. 0.15** *Planck* CMB map (as rendered by SMICA, see text).

The two other algorithms are rather based on minimising the variance of the CMB component. The first one, NILC, is an internal linear combination (ILC) working in needlet (wavelet) domain. It makes an ILC at each needlet scale independently. It uses the 44 to 857 GHz channels and yields a $5^t$ resolution map. Finally, SEVEM minimises the CMB variance by template fitting. This method operates in the pixel domain and uses internal templates which are produced by subtracting adjacent frequency channels after smoothing them to a common resolution; these 4 templates correspond to the difference maps (30 - 44), (44 - 70), (857 - 535), (545 - 353). They are used to clean the 143 and 217 GHz maps, which can optionally be combined afterwards in harmonic space to produce a single CMB map at $5^t$ resolution.

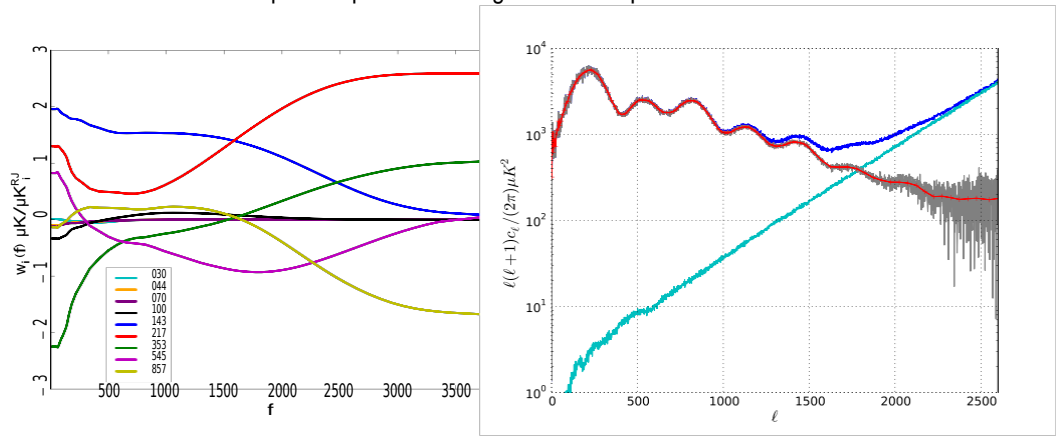

**Fig. 0.16** *Planck* CMB map as rendered by SMICA. a) Weights of the harmonics of the various frequency channels to obtain the harmonics of the CMB map (see text). b) SMICA map spectra.

Tests performed on simulations lead us to elect the SMICA CMB map of Fig. 0.15 as the reference case, but with each of the others offering unique capabilities of their own. Figure 0.16-a) illustrates concretely the trade-off performed by the SMICA method to minimize the sum of noise and foregrounds residuals at each scale, given the available input channels which each have their own noise level and effective beam. The figure shows that most of the positive weight to estimate the CMB is given to the 143 GHz channel at low-$\ell$ and to the 217 GHz one at higher-$\ell$. This is not too surprising since the 143 GHz channel is the most sensitive (less noisy) channel of *Planck*, but with a resolution limited to 7$^t$; at higher angular resolution, the next most sensitive channel is the 217 GHz one, which is endowed with a 5$^t$ angular resolution. Of course, the data combination must also null the foregrounds contribution, and the dominant negative weights are those of the 353 GHz at low-$\ell$, the 545 GHz at mid-$\ell$ and 857 GHz at larger-$\ell$. This is also not unreasonable, since at 143 and 217 GHz, the dominant foreground contribution comes from the dust emission of the Galaxy, which is dominating the sky emission at all the highest frequencies (350-857 GHz). All other channels are clearly subdominant on all scales and are likely used to finely cancel further foreground contributions, principally at low-$\ell$. Figure 0.16-b) shows the angular power spectrum of the resulting map in dark blue, as well as an estimate in light blue of its noise. The latter was obtained by processing with the same weights the half-difference of the half-ring maps at each frequency (which each provides a good estimate of the noise per frequency channel). The red curve shows the noise-de-biased power spectrum, with a flattening at hight-$\ell$ betraying the presence of residual point source emission, which a detailed power spectrum analysis confirms. The plot also shows that this CMB map is nearly noise- and foreground-free till $\ell = 1500$, a range where most of the cosmological information lies in the case of the $\Lambda$CDM model.

We will return later to the direct scientific exploitation of the CMB map to address the question of possible deviations from a Gaussian stationary field. For now we turn to the characterisation of its two-point statistics, which is exhaustive in the Gaussian case.

### 0.3.2 *Planck* CMB spectra and likelihood

The straightforward way to proceed in order to determine the extent to which a given theoretical angular power spectrum, $C_\ell$, is a good match to the *Planck* determination of the CMB spatial distribution is to use a pixel-based maximum likelihood approach. If $\mathbf{m}$ is a vector gathering all $n$ pixel values of an empirical CMB map, let us assume that it is simply a superposition of two other maps, that of the true CMB sky, $\mathbf{s}$, and a noise map, $\mathbf{n}$:

$$\mathbf{m} = \mathbf{s} + \mathbf{n}. \qquad (0.1)$$

If we further assume that the Gaussian approximation holds to describe the anisotropies and the noise, then we can assess the probability of measuring the map given the $n \times n$ covariance matrix, $\mathbf{M}$, from which it is drawn,

$$P(\mathbf{m}|C_\ell) = \frac{1}{2\pi^{n/2}|\mathbf{M}|^{1/2}} \exp\left(-\frac{1}{2}\mathbf{m}^t \mathbf{M}^{-1} \mathbf{m}\right), \qquad (0.2)$$

Bayes theorem states that the likelihood of a given theoretical $C_\ell$, $\mathrm{L}(C_\ell)$, is given (up to a normalization factor irrelevant here) by $\mathrm{L}(C_\ell) \propto P(\mathbf{m}|C_\ell)P(C_\ell)$, where $P(C_\ell)$ encode the prior knowledge we had on ($C_f$) before incorporating the new map measurement. The data covariance matrix, **M**, is given by:

$$M(C_f) = C(C_f) + N.$$

The pixel-space noise covariance matrix, **N**, depends on the details of how the map was obtained (e.g. the scanning strategy and noise properties in the time domain, the relative weights of the detectors); it needs to be provided by the team who derived that map. The CMB map covariance matrix between pixels, $C = \langle s^T s \rangle$, depends on the theoretical angular power spectrum in harmonic space, $C_f$, as

$$\langle s_{i_1} s_{i_2} \rangle = \sum_{f=2}^{f_{max}} \frac{2f+1}{4\pi} \hat{C}_f P_f(\theta_{i_1 i_2}) + N_{i_1 i_2}, \quad \text{with} \quad \hat{C}_f = C_f b_f^2 w_f^2, \tag{0.3}$$

where ($i_1, i_2$) are pixel indices, $b$ is the beam window function and $w$ is the pixel window function of the map. Estimating the likelihood of a theoretical $C_f$, given a map and its characteristics (noise, resolution), is therefore quite straightforward by using the above, albeit with a little catch: a numerical evaluation of such a likelihood requires $O(n^3)$ operations, with $n \approx 5 \times 10^7$ in the *Planck* case. For such a large $n$, this can only be done in practice for $f_{max} \lesssim 30$, which is a bit restrictive, as compared to the range $f_{max} \gtrsim 2500$ were *Planck* has still some constraining power.

This snag can be circumvented by an hybrid likelihood approach where at large scales (low-$f$) one uses a modification using maps which we shall now describe, and at small scales (high-$f$) a Gaussian likelihood approximation on power spectra. At low-$f$, in order to account for the impact of the existence of foregrounds, we use the Commander solution where the foregrounds are parametrised at the map level and these parameters are marginalised over using Gibbs samples (*i.e.*, each parameter of the model is updated in turn). Let us start by generalising the sky model to include $n_f$ foregrounds and consider $n_v$ frequency maps:

$$d_v = s + \sum_i f_v^i + n_v. \tag{0.4}$$

One can map out the posterior distribution $P(s, f, C_f | d)$ by Gibbs sampling using sequentially:

$$\text{a multivariate Gaussian conditional distribution,} \quad P(s | f, C_f, d) \rightarrow s, \tag{0.5}$$
$$\text{a conditional obtainable numerically,} \quad P(f | s, C_f, d) \rightarrow f, \tag{0.6}$$
$$\text{an inverse gamma distribution,} \quad P(C_f | s, f, d) \rightarrow C_f, \tag{0.7}$$

which allows generating many realisations of the CMB sky. From *each* realisation, $s_k$, we can derive a likelihood function,

$$L^k(C_f) \propto \frac{\sigma_{f,k}^{\frac{2f-1}{2}}}{C_f^{\frac{2f+1}{2}}} \exp\left(-\frac{2f+1}{2}\frac{\sigma_{f,k}}{C_f}\right), \quad \text{with} \quad \sigma_{f,k} \equiv \frac{1}{2f+1}\sum_{f=-m}^{n} |a_{fm}^k|^2. \tag{0.8}$$

This likelihood is only correct though in the absence of noise, of foregrounds, and of a sky mask. But one can show that all these are accounted for by using the the Blackwell-Rao estimate:

$$L(C_{\mathbf{f}}) \propto \sum_{k=1}^{n_{samp}} L^k(C_{\mathbf{f}}), \qquad (0.9)$$

which properly include the uncertainties they induce.

The Commander implementation of this pixel-based approach can currently be used till $f \sim 60$ (although *Planck* 2013 analysis relies on it only in the multipole interval [2, 49]), and we employed the following signal model parametrisation: one CMB map, a single low-frequency Galactic component (2 maps: an amplitude and a power law index for the emission law), one map of the CO emission (with fixed line ratio at different frequencies), one single dust component (2 maps). The foreground intensity and spectral parameters maps are marginalised over by Gibbs sampling, and one derives Blackwell-Rao estimates of the posterior on $C(\mathbf{f})$ at individual multipoles.

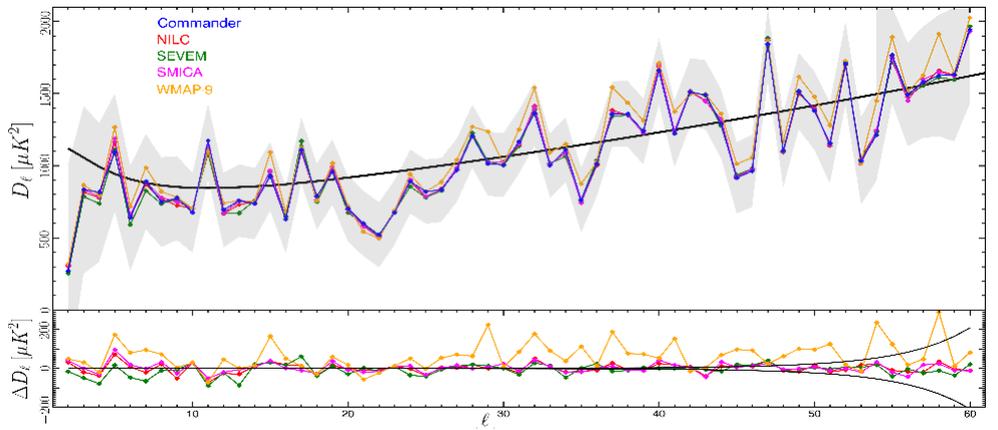

**Fig. 0.17** CMB temperature anisotropies angular power spectrum at low-$f$, as estimated with Commander, NILC, SEVEM, or SMICA, and the 9-year WMAP ILC map, using the Bolpol quadratic estimator. The grey shaded area indicates the 1 σ Fisher errors while the solid line shows the Planck ΛCDM best fit model. Bottom panel: Power spectrum differences for each algorithm/data set, relative to the Commander spectrum, as estimated from the spectra shown in the panel above. The black lines show the expected 1 σ uncertainty due to (regularization) noise.

Figure 0.17 compares the resulting Commander estimate with empirical spectra from our other CMB cleaning methods, and the latest and last WMAP determination. *Planck* results match each other very well, but their comparison with that of WMAP clearly shows a systematic difference, which can be described as a $\sim 1.3\%$ calibration difference, which is currently under investigation by both teams (and whose influence for cosmology has been carefully bounded). Otherwise, the agreement is quite good, and in any case, the differences, while significant, are much lower than the intrinsic cosmic variance limiting the cosmological analysis.

We have also developed two component separation methods, CamSpec & Plik, specifically dedicated to the spectrum determination of the CMB at high-$f$. They are both directly based on power spectra, starting from mask- and beam-deconvolved empirical power spectra ("à la Master / Spice"), for all 56 possible detector-based map pairs in the 100-217 GHz range.

They both account for the (small) uncertainties in the relative calibration of the detector sets used, as well as that in the determination of their window function, but they differ in their handling of the correlation between scales induced by the masks and their use of detectors sets at the same frequencies[10]. We designed parametrised foreground power spectrum templates which can be used to analyse jointly *Planck* with the ACT & SPT likelihoods. These foreground templates depend on 11 (or 13) parameters[11]. Fig. 0.18 shows one fit solution using all detector spectra on about 40% of the sky versus the individual components of the model and their sum. One clearly sees that, once a sufficient region has been masked (using a threshold on the intensity of the emission at 350 GHz), the Galactic dust emission does not contribute much, while the Poisson and clustered source contributions of compact source are the dominant high-$\ell$ contributions, reaching values similar to that of the primary CMB at $\ell \sim 2500$. We note too that at $\ell \lesssim 1500$ where most of the information lies in the (standard, no extensions) $\Lambda$CDM model, both foregrounds and noise are quite low in that range.

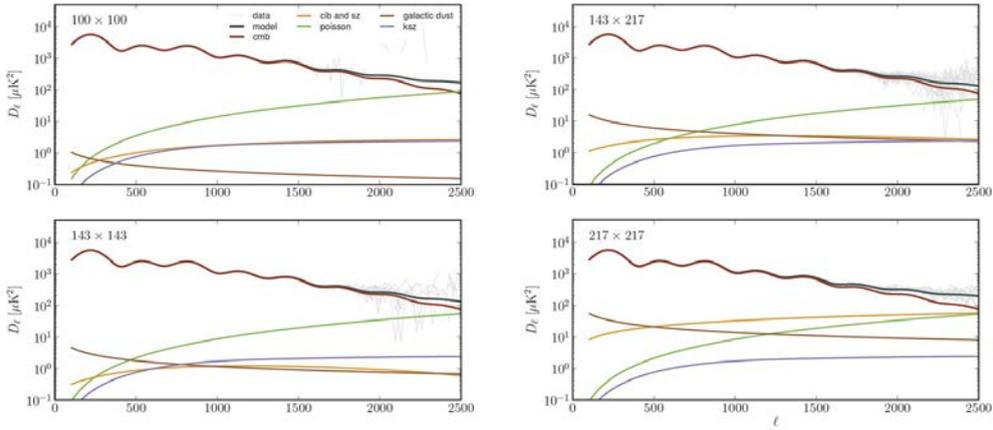

**Fig. 0.18** Decomposition of the total best-fitting model power spectra into CMB (red), combined thermal SZ and CIB (green), unresolved point sources (green), kinetic SZ, and Galactic dust, using the Plik code. The light grey lines show the individual detector pairs within each frequency combination.

The (2013) high-$\ell$ reference likelihood, based on CamSpec, relies on a quite conservative data selection in order to minimise the impact of foregrounds and thus of any (small) inaccuracies in their modelling. We thus kept only the easiest to model & most informative frequency maps, masking variable portions of them in order to simplify the modelling while still retaining as much sky as possible (to reduce cosmic variance). We also tailored the multipole range used in each case. Fig. 0.19 illustrates the frequency and ℓ-range aspect of our data selection. In addition, to minimise cosmic variance, we have used a smaller mask at 100 GHz, retaining $f_{sky} = 49\%$ of the sky[12] when dust emission is quite weak, while we have been more stringent for the 143, 143 × 217 and 217 GHz cross-spectra which have been obtained on about 30 % of the sky (we do not use the 100 × 143 and 100 × 217 spectra which would be delicate to model accurately). Note that the empirical spectra used by CamSpec have also been corrected for dust emission by using in a pre-processing step a 850 GHz-based template. Fig.0.20 shows the spectra used in the likelihood, after removing our (common) foreground solution (top) and our common CMB best fit model (bottom). This comparison is clear evidence of the high degree of consistency of these spectra with our modelling.

---

[10] In CamSpec, after careful analysis of the consistency of all the individual spectra within one frequency pair, the data has been optimally compressed into combined cross-spectra (allowing for small recalibration within a frequency pair, well within the prior derived from calibration accuracy analyses, and computing the beam uncertainties of that combination).

[11] We use 4 parameters for describing the (clustered) Cosmic Infrared Background component (CIB), 4 other parameters for the levels of the Poisson fluctuations from unresolved sources in each frequency combination, 2 more parameters for the amplitude of the Sunyaev-Zeldovich effect and 1 for its cross-correlation with the CIB. In the Plik case, we also have two additional parameters to describe the dust emission since this contribution is not reduced by a pre-processing step as in CamSpec.

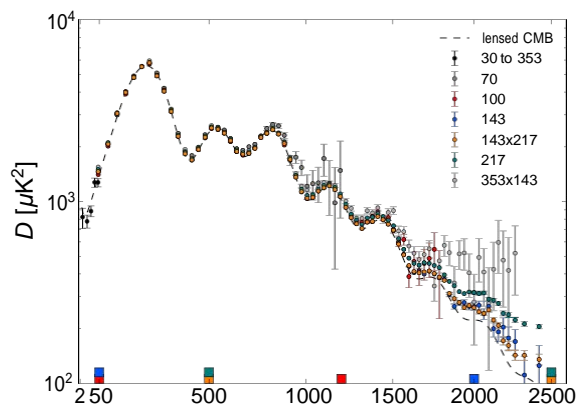

**Fig. 0.19** *Planck* power spectra and data selection. The coloured tick marks indicate the ℓ-range of the four cross-spectra included in CamSpec (here all computed with the same mask). Although not used, the 70 GHz and 143 × 353 GHz spectra demonstrate the consistency of the data. The dashed line indicates the best-fit Planck spectrum. The black points show the low-ℓ points from Commander.

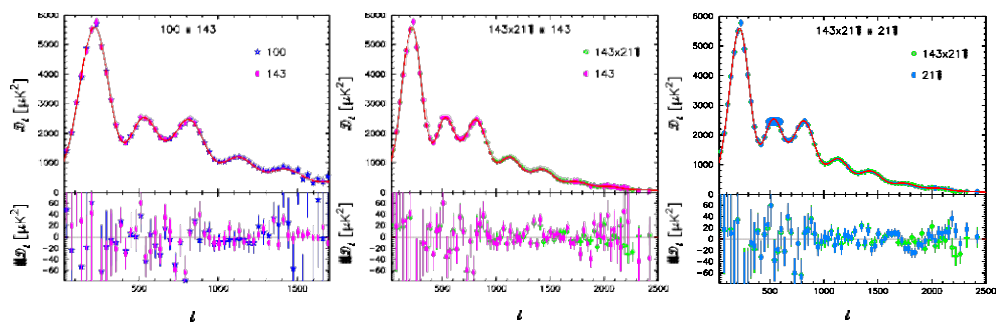

**Fig. 0.20** Comparison of pairs of foreground subtracted cross spectra, demonstrating consistency of the residuals with respect to the best-fit theoretical model. The red line in each of the upper panels shows our best fit six parameter ΛCDM spectrum. The lower panels show the residuals with respect to this spectrum, together with error bars computed from the diagonal components of the covariance matrices of the band averages. The points here are band-averaged in bins of width Δℓ ∼ 31.

---

[12] This is the sky fraction remaining after excising areas where point sources have been resolved and included in our point sources catalogues, in addition to removing the part with too strong Galactic emission, as traced by the 350 GHz Intensity..

Our CamSpec high-$\ell$ likelihood[13] is approximated by a Gaussian,

$$p = \exp(-S), \quad \text{with} \quad S = \frac{1}{2}(\hat{X} - X)^T \hat{M}^{-1}(\hat{X} - X), \qquad (0.10)$$

where $\hat{X}$ stands for the data vector ($\hat{C}_f^{100\times100}$, $\hat{C}_f^{143\times143}$, $\hat{C}_f^{217\times217}$, $\hat{C}_f^{143\times217}$). The covariance matrix of these frequency spectra spectra, $\hat{M}$, is computed for a fixed fiducial model, which includes a model for the CMB, the noise - both correlated and anisotropic, and the foregrounds
co
a
lil
ex

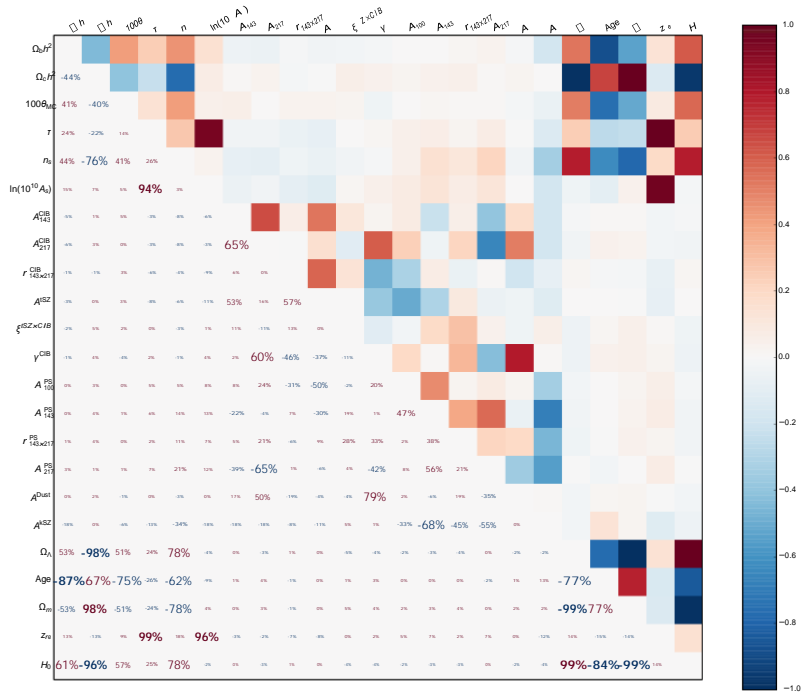

**Fig. 0.21** Correlation matrix between all the cosmological (top block), foreground (middle block), and derived (bottom block) parameters, estimated by sampling the Plik likelihood.

Figure 0.21 provides direct evidence of the lack of sensitivity of the cosmological parameters derived with *Planck* high-$\ell$ data to the foreground parameters by displaying their cor-relations. One first sees the well-known correlations between cosmological parameters, both primary and derived ones. One also sees the important correlation between the foreground parameters, since the dynamical range of *Planck* is not large enough to fully break some of the degeneracies stemming from the flexibility of our foregrounds modelling. But it is rather comforting to note the weakness of the correlations between the cosmological and foreground parameters, illustrating the fact that *Planck* does constrain well the sum of the foregrounds contribution per frequency, allowing a robust CMB separation.

---

[13] Our Plik alternative at high-$\ell$ relies on a different approximation, the Kullback divergence between the empirical and model spectra; we use it for cross-check and robustness tests. Indeed it is faster, but requires to work on binned data to minimize the correlation between $\ell$'s which the method assumes negligible.

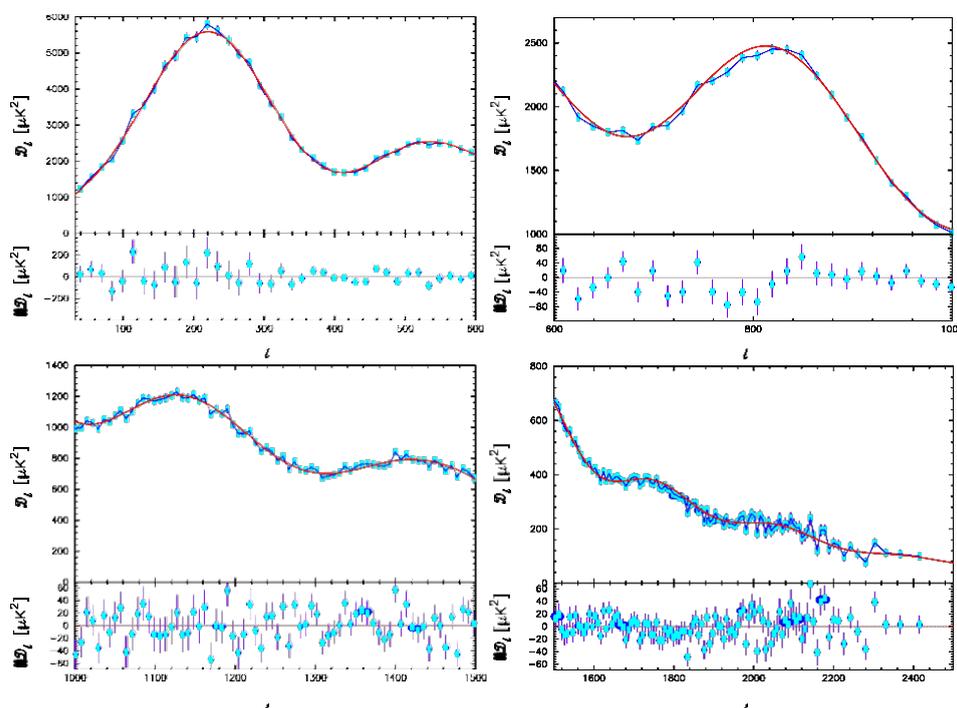

**Fig. 0.22** Zoom-in of regions of the *Planck* high-$\ell$ CMB power spectrum using fine bin widths ($\Delta\ell = 15$ for $\ell < 1000$ and $\Delta\ell = 7$ for $1000 \leq \ell \leq 2200$. In the upper panels, the red lines show the best-fit $\Lambda$CDM spectrum, and the blue lines join the *Planck* data points. The error bars are computed from the diagonal elements of the band-averaged covariance matrix, including contributions from foreground and beam transfer function errors.

Figure 0.22 allows visualising details of our high-$\ell$ CMB power spectrum with small bins (top panel, obtained by removing the best fit foreground model), and residuals with respect to our $\Lambda$CDM best fit (in the bottom panel). One should note that the correlated fluctuations seen in this figure are mask-induced, and perfectly compatible with the six parameter $\Lambda$CDM model. Features such as the 'bite' missing from the third peak at $\ell \sim 800$ or the oscillatory features in the range $1300 \lesssim \ell \lesssim 1500$ are not unexpected on the basis of our covariance matrix, as we checked on simulations[14].

---

[14]Figure 0.20 also hints at a small bite around $\ell = 1800$ in the 217 GHz spectrum, which we found, after the data release, to be not statistical but rather due to a residual of an EM interference line. We estimate in an annex of the

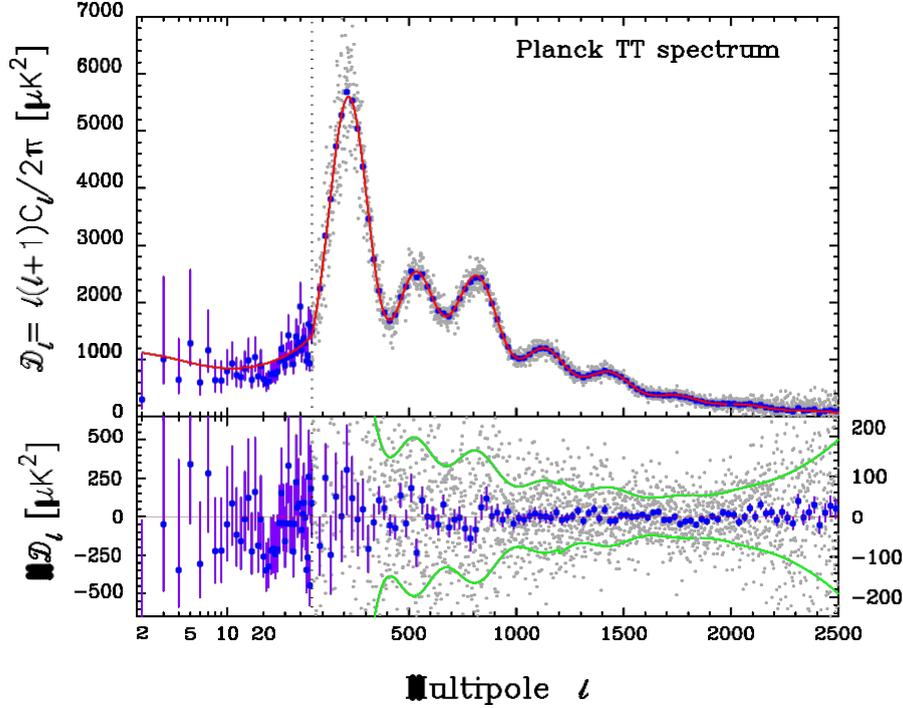

**Fig. 0.23** *Planck* foreground-subtracted temperature power spectrum, with foreground and other "nuisance" parameters fixed to their best-fit values for the base ΛCDM model. The red line shows the temperature spectrum for the best-fit base ΛCDM cosmology. The lower panel shows the power spectrum residuals with respect to this theoretical model. The green lines show the ±1 σ errors on the individual power spectrum estimates at high multipoles. In the lower panel, note the change in vertical scale in the left & right axis at $\ell = 50$.

The full *Planck* CMB likelihood is given by the product of our low-$\ell$ pixel-based likelihood and of high-$\ell$ spectra-based, with a sharp transition at $\ell = 50$ (we did checked the weakness of the correlation between the $C_\ell$ across that $\ell$-value). The resulting foreground-subtracted temperature power spectrum (with foreground and other "nuisance" parameters fixed to their best-fit values for the base ΛCDM model) is displayed in Fig. 0.23.

In summary, the power spectrum at low multipoles ($\ell = 2$–49, plotted on a logarithmic multipole scale in Fig. 0.23) is determined by the Commander algorithm applied to the *Planck* maps in the frequency range 30–353 GHz over 91% of the sky. This is used to construct a low-multipole temperature likelihood using a Blackwell-Rao estimator. The asymmetric error bars of the figure show 68% confidence limits and include the contribution from uncertainties in foreground subtraction. At multipoles $50 \leq \ell \leq 2500$ (plotted on a linear multipole scale) we show the best-fit CMB spectrum computed from the CamSpec likelihood after removal of unresolved foreground components. The light grey points show the power spectrum multipole-by-multipole. The blue points show averages in bands of width $\Delta\ell \approx 31$ together parameter paper that it can shift the best fit ΛCDM value by at most a small fraction of a σ, at a level comparable with other uncertainties from, e.g. foreground modelling.

with 1 σ errors computed from the diagonal components of the band-averaged covariance matrix (which includes contributions from beam and foreground uncertainties). The red line shows the temperature spectrum for the best-fit base ΛCDM cosmology. The lower panel shows the power spectrum residuals with respect to this theoretical model. The green lines show the ±1 σ errors on the individual power spectrum estimates at high multipoles computed from the CamSpec covariance matrix.

Since, as described earlier, the 2013 *Planck* data release does not include polarisation, this data can only break weakly the $A_s - \tau$ degeneracy. But *Planck* data also allows a determination of the lensing of the CMB by the large scale structures traversed by the CMB photons on their path to the observer. This allows to (partially) lift the $A_s - \tau$ degeneracy, and the next section is devoted to the determination of the likelihood of the lensing potential power spectrum (which will allow to further constrain the parameters which controls it) . Finally, let us note that we have also included the possibility to optionally use WMAP polarisation data (assuming negligible noise in TT, TQ, TU, and vanishing B modes) by using WMAP 9 years polarisation likelihood, but with *Planck* determination of the CMB temperature distribution to compute the TE correlations. For the convenience of users of Planck likelihood, we completed the package with the likelihoods of ACT and SPT, which we can be analysed jointly, since we built from start a consistent model to allow using them together.

### 0.3.3 *Planck* CMB lensing spectrum and likelihood

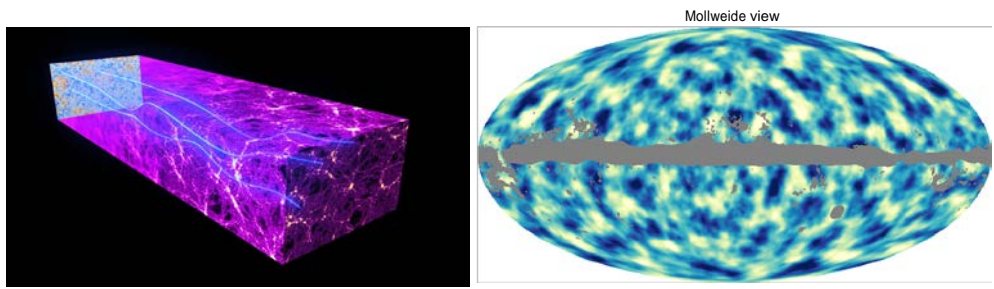

**Fig. 0.24** a) Artist view (not to scale) illustrating how CMB photons from the last scaterring surface are deflected by the gravitational lensing effect of massive cosmic structures as they travel across the Universe. As a result, the initial pattern is deformed, in a specific way which betrays the intervening mass distribution. Copyright: ESA and the Planck Collaboration. b) Wiener-filtered lensing potential maps in Galactic coordinates (Mollweide projection). This is an extended reconstruction ($f_{sky}$ = 0.78) based on the NILC CMB map.

As first considered by Blanchard and Schneider (Blanchard and Schneider, 1987), the large-scale structures of the Universe which intercede between ourselves and the CMB last-scattering surface induce small but coherent (Cole and Efstathiou, 1989) deflections of the observed CMB temperature and polarisation anisotropies, with a typical magnitude of $2^t$ (see Fig. 0.24-a)). These deflections blur the acoustic peaks (Seljak, 1996), generate small-scale power (Linder, 1990; Metcalf and Silk, 1997), non-Gaussianity (Bernardeau, 1997), and convert a portion of the dominant *E*-mode polarisation to *B*-mode (Zaldarriaga and Seljak, 1998). This gravitational lensing of the CMB therefore obscures the primordial fluctuations (Knox and Song, 2002), but it also provide a measure of the distribution of mass in the Universe at intermediate redshifts (typically $0.1 < z < 5$). *In short, lensing introduces 2-3' deflections, coherent over 2-3 degrees, mainly coming from redshifts of 2-3!*

In the $\Lambda$CDM framework, there exist accurate methods to calculate the effects of lensing on the CMB power spectra (Challinor and Lewis, 2005), as well as optimal estimators for the distinct statistical signatures of lensing (Hu and Okamoto, 2002; Hirata and Seljak, 2003a). Since lensing performs a remapping of the CMB fluctuations, the observed temperature anisotropy in direction $\hat{n}$ is given in terms of the unlensed, "primordial" temperature anisotropy, $T$, as

$$\tilde{T}(\hat{n}) = T(\hat{n} + \nabla \varphi(\hat{n})),$$
$$= T(\hat{n}) + \nabla \varphi(\hat{n}) \cdot \nabla T(\hat{n}) + O(\varphi^2), \tag{0.11}$$

where $\varphi(\hat{n})$ is the CMB lensing potential, a line-of-sight integral[15] of the gravitational potential. This lensing potential is a measure of the integrated mass distribution back to the last-scattering surface. It is affected by effects that affect distance scales and the growth rate of structure in the late Universe. To first order, its effect on the CMB is to introduce a correlation between the lensed temperature and the gradient of the unlensed temperature, a property which can be exploited to make a (noisy) reconstruction of the lensing potential itself (Okamoto and Hu, 2003; Hirata and Seljak, 2003b), of the general form

$$\bar{\varphi} = \mathcal{N}^{-1} \nabla \cdot [C^{-1} T \nabla (C^{-1} T)]. \tag{0.13}$$

This quadratic estimator is a weighted sum of the product of a filtered version of the temperature map[16] and of its derivative, which can be made optimal with a proper choice of weight functions and filtering. Figure 0.24-b) displays a lensing potential map obtained by *Planck* (in this case based on the NILC CMB map).

It is of great interest to construct the angular power spectrum of the lensing map, $C_L^{\varphi\varphi}$, which therefore probes the lensing-induced 4-point correlator, albeit biased by the presence of (reconstruction) noise. Figure 0.25-a) displays the level of this bias using different (masked) channel maps as input temperature map, in comparison to the expected (dashes) lensing spectrum predicted in the $\Lambda$CDM model for *Planck* best fit parameters. It shows that the best signal-to-noise ratio per individual mode is only three quarters around the multipole $L \sim 40$ (around 4 degrees), even with a minimum variance combination from the 3 best CMB channel maps. The main limitation arises from a relatively poor resolution of the *Planck* CMB map (limiting the number of modes to observe/average over) rather than noise or foregrounds residuals.

---

[15] The lensing potential $\varphi(\hat{n})$ is defined by

$$\varphi(\hat{n}) = -2 \int_0^{\chi_*} d\chi \frac{f_K(\chi_* - \chi)}{f_K(\chi_*) f_K(\chi)} \Psi(\chi \hat{n}, \eta_0 - \chi). \tag{0.12}$$

Here $\chi$ is conformal distance (with $\chi_* \approx 14000$ Mpc) denoting the distance to the CMB last-scattering surface), $f_K(\chi)$ is the angular-diameter distance which depends on the curvature of the Universe, $K$, and $\Psi(\chi \hat{n}, \eta)$ is the gravitational potential at conformal distance $\chi$ along the direction $\hat{n}$ at conformal time $\eta$ (the conformal time today is denoted as $\eta_0$).

[16] The filtered map multipoles, $\bar{T}_{\ell m}$ are given by $\bar{T}_{\ell m} = (C^{-1} T)_{\ell m}$, where $T$ is a beam deconvolved CMB map and $C$ is its total signal+noise covariance matrix.

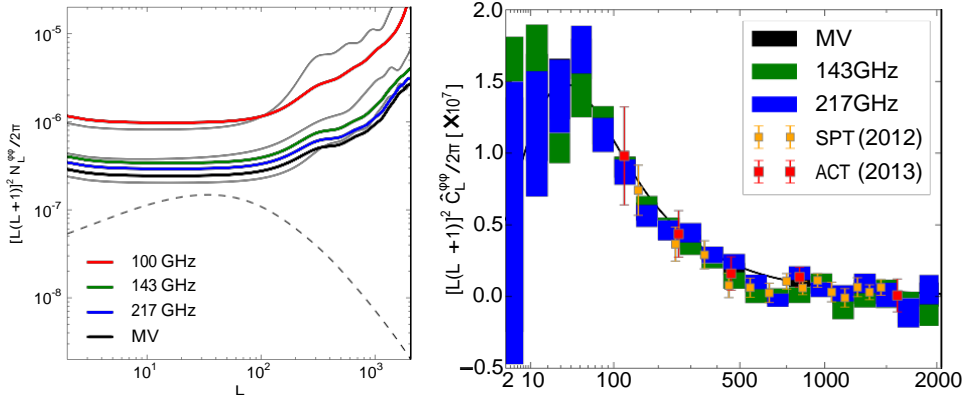

**Fig. 0.25** Lensing potential spectrum. a) Sky-averaged lens reconstruction noise levels for the 100, 143, and 217 GHz *Planck* channels (red, green, and blue solid, respectively), as well as for experiments that are cosmic-variance limited to a maximum multipole $f_{max}$ = 1000, 1500, and 1750 (upper to lower solid grey lines). The dashed black line shows the expected level from *Planck* temperature best fit model. The noise level for a minimum-variance ("MV") combination of 143+217 GHz is shown in black (the gain from adding 100 GHz is negligible). b) *Planck* measurements of the lensing power spectrum compared to the prediction for the best-fitting *Planck*+WP+highL $\Lambda$CDM model parameters, and to the SPT and ACT bandpowers . The plot shows the ±1 $\sigma$ error from the diagonal of the covariance matrix. All three experiments are consistent between them and with the $\Lambda$CDM prediction, which itself has quite small uncertainties due to the uncertainties of the $\Lambda$CDM parameters).

The determination of the lensing potential spectrum therefore requires a large "noise" debiasing, which requires great care in its evaluation, and whose uncertainty contributes to the uncertainty of the spectrum itself. Once this is done, fig. 0.25-b) shows the striking agreement of the resulting (binned) spectra of our individual 143 and 217 GHz reconstructions as well as their minimum variance combination. They are in reasonable agreement with the expectation from our best-fit $\Lambda$CDM model[17]. The *Planck* determination extends but is quite compatible with the determinations from the ACT and SPT experiments (Das *et al.*, 2013; van Engelen *et al.*, 2012).

Based on our measurements of the lensing potential power spectrum in the $40 \leq L \leq 400$ range, we constructed a Gaussian likelihood based on (bins in) $C_L^{\varphi\varphi}$ of the form

$$-2 \ln L_\varphi(C^{\varphi\varphi}) = \hat{C}^{\varphi\varphi} - C^{\varphi\varphi} \Sigma^{-1} \hat{C}^{\varphi\varphi} - C^{\varphi\varphi}, \quad (0.14)$$

where $\hat{C}^{\varphi\varphi}$ is our data vector and $\Sigma$ is the covariance matrix (between bins). The paper (Planck Collaboration XVII, 2013) develops further the use of the lensing map (cross-correlations with the internally determined CIB signal, with the ISW effect, and with external probes), and provides a detailed account of that likelihood usage (e.g. a factor two improvement on the curvature-Dark energy constraints of CMB likelihood alone), which we will only use below in combination with the CMB likelihood.

---

[17] Dividing the L multipole range [1, 2048] into bins of $\Delta L$ = 64 and binning uniformly in $L(L+1)^2 C_L^{\varphi\varphi}$, we obtain a reduced $\chi^2$ for the difference between our power spectrum estimate and the model of 40.7 with 32 degrees of freedom. The associated probability to exceed is 14%. Furthermore, the mere detection of the amplitude of our fiducial model corresponds to a 25 $\sigma$ detection of gravitational lensing effects.

In summary, our baseline lensing likelihood is based on an $f_{sky}$ ~70% minimum variance combination of the 143 and 217 GHz frequency maps, after dust correction using the 857 GHz map as a template. The robustness against foreground contamination was assessed by comparing to results from properly cleaned CMB maps using component separation technique. The foreground-cleaned maps were used to obtained a robust lensing reconstruction on ~ 90% of the sky. And the 2014 data release, with about twice more data should bring a ~ 25% decrease of the $C_f^{\phi\phi}$ uncertainties (and further investigation of possible systematics, attempts at reducing the level of conservatism of some choices, inclusion of polarization...).

## 0.4 CMB cosmological consequences (some)

The definitions of all the cosmological parameters we use are given in Table 0.2, together with priors and default values. Detailed description may be found in the *Planck* "Cosmological parameters" article (Planck Collaboration XVI, 2013).

### 0.4.1 The base ΛCDM model

Using the *Planck* only likelihood introduced above (temperature+lensing), we find a good match with a basic ΛCDM model with a minimal set of six parameters ($\omega_b$, $\omega_c$, $\theta$, $\tau$, $n_s$, $A_s$) which will refer to as our *base model* in all the following. Figure 0.26 allows comparing constraints on pair of these parameters, as well as their posterior marginals, in case of using *Planck* alone (CMB+lensing, colour-coded samples), or adding *WMAP* polarisation information (to further break the $A_s - \tau$ degeneracy, red contours), both of which are compatible but more precise than the *WMAP*-9 alone constraints (grey contours). This figure illustrates well the level of consistency of parameters from *Planck* alone with those inferred from *WMAP*-9 alone, as well as the narrowing of their allowed range induced by the additional information brought by *Planck*.

Table 0.3 displays the *Planck* (CMB+lensing) constraints on the base ΛCDM model in numerical form. Let us only note for now that the angular size of the sound horizon when the optical depth is unity, $\theta_*$, is determined with an accuracy of 0.06% (1 $\sigma$) thanks to to the excellent determination of the peaks separation allowed by measuring precisely the first 7 peaks of the angular power spectrum in a single experiment. It is also worth noting that the 68% limit on the power-law index of the scalar spectrum, $n_s$, by *Planck* alone is 0.9635 ± 0.0094 (0.9% 1 $\sigma$ accuracy), *i.e.*, exact scale invariance is already excluded at the 3.9 $\sigma$ level. Regarding derived parameters, let us simply note for now that *Planck*+lensing leads to a low value of $H_0 = 67.9 \pm 1.5$ km s$^{-1}$ Mpc$^{-1}$, a relatively low matter density $\Omega_m = 0.307 \pm 0.019$, a slightly older Universe with an age of 13.796 ± 0.058 Gyr and a primordial Helium abundance, $Y_P = 0.24775 \pm 0.00014$ (0.05% !), fully consistent with BBN constraints.

As mentioned earlier, the foreground model used in the *Planck* high-**f** likelihood was designed to allow a joint analysis with the CMB likelihoods from the two currently leading small scale experiments ACT & SPT. Figure 0.27 shows the *Planck* determination of the CMB temperature spectrum, together with those from *WMAP*-9, ACT and SPT, demonstrating a spectacularly precise and consistent determination of the spectrum of the temperature anisotropies over 3 decades in scale and nearly three in amplitude[18]. Concerning the parameters, fig. 0.28 allows comparing the constraints on the six ΛCDM parameters from these various data combinations. It is first interesting to confirm that the inclusion of *WMAP* polarisation mainly breaks the $A_s - \tau$ degeneracy of *Planck* CMB alone (it does it better than *Planck* temperature lensing data), while barely modifying the constraints from other parameters. We also see that adding the high-$\ell$ data from SPT & ACT brings very little change to the base parameters, which is not surprising either since most of the information in this model can be shown to lie at $\ell \lesssim 1800$, where *Planck* alone is already very close to a cosmic variance limited experiment for measuring temperature anisotropies.

**Table 0.2** Cosmological parameters used in our analysis. For each, we give the symbol, prior range, value taken in the base $\Lambda$CDM cosmology (where appropriate), and summary definition. The top block contains parameters with uniform priors that are varied in the MCMC chains. The ranges of these priors are listed in square brackets. The lower blocks define various derived parameters.

| Parameter | Prior range | Baseline | Definition |
|---|---|---|---|
| $\omega_b \equiv \Omega_b h^2$ | [0.005, 0.1] | ... | Baryon density today |
| $\omega_c \equiv \Omega_c h^2$ | [0.001, 0.99] | ... | Cold dark matter density today |
| $100\theta_{MC}$ | [0.5, 10.0] | ... | 100 × approximation to $r_*/D_A$ (CosmoMC) |
| $\tau$ | [0.01, 0.8] | ... | Thomson scattering optical depth due to reionization |
| $\Omega_K$ | [−0.3, 0.3] | 0 | Curvature parameter today with $\Omega_{tot} = 1 - \Omega_K$ |
| $m_\nu$ | [0, 5] | 0.06 | The sum of neutrino masses in eV |
| $w_0$ | [−3.0, −0.3] | −1 | Dark energy equation of state[a], $w(a) = w_0 + (1 - a)w_a$ |
| $w_a$ | [−2, 2] | 0 | As above (perturbations modelled using PPF) |
| $N_{eff}$ | [0.05, 10.0] | 3.046 | Effective number of neutrino-like relativistic degrees of freedom |
| $Y_P$ | [0.1, 0.5] | BBN | Fraction of baryonic mass in helium |
| $A_L$ | [0, 10] | 1 | Amplitude of the lensing power relative to the physical value |
| $n_s$ | [0.9, 1.1] | ... | Scalar spectrum power-law index ($k_0 = 0.05 \mathrm{Mpc}^{-1}$) |
| $n_t$ | $n_t = -r_{0.05}/8$ | Inflation | Tensor spectrum power-law index ($k_0 = 0.05 \mathrm{Mpc}^{-1}$) |
| $dn_s/d\ln k$ | [−1, 1] | 0 | Running of the spectral index |
| $\ln(10^{10} A_s)$ | [2.7, 4.0] | ... | Log power of the primordial curvature perturbations ($k_0 = 0.05 \mathrm{Mpc}^{-1}$) |
| $r_{0.05}$ | [0, 2] | 0 | Ratio of tensor primordial power to curvature power at $k_0 = 0.05 \mathrm{Mpc}^{-1}$ |
| $\Omega_\Lambda$ | | ... | Dark energy density divided by the critical density today |
| $t_0$ | | ... | Age of the Universe today (in Gyr) |
| $\Omega_m$ | | ... | Matter density (inc. massive neutrinos) today divided by the critical density |
| $\sigma_8$ | | ... | RMS matter fluctuations today in linear theory |
| $z_{re}$ | | ... | Redshift at which Universe is half reionized |
| $H_0$ | [20,100] | ... | Current expansion rate in km s$^{-1}$ Mpc$^{-1}$ |
| $r_{0.002}$ | | 0 | Ratio of tensor primordial power to curvature power at $k_0 = 0.002 \mathrm{Mpc}^{-1}$ |
| $10^9 A_s$ | | ... | $10^9 \times$ dimensionless curvature power spectrum at $k_0 = 0.05 \mathrm{Mpc}^{-1}$ |
| $\omega_m \equiv \Omega_m h^2$ | | ... | Total matter density today (inc. massive neutrinos) |
| $Y_P$ | | bbn | Fraction of baryonic mass in Helium |
| $z_*$ | | ... | Redshift for which the optical depth equals unity (see text) |
| $r_* = r_s(z_*)$ | | ... | Comoving size of the sound horizon at $z = z_*$ |
| $100\theta_*$ | | ... | 100 × angular size of sound horizon at $z = z_*$ ($r_*/D_A$) |
| $z_{drag}$ | | ... | Redshift at which baryon-drag optical depth equals unity |
| $r_{drag} = r_s(z_{drag})$ | | ... | Comoving size of the sound horizon at $z = z_{drag}$ |
| $r_{drag}/D_V(0.57)$ | | ... | BAO distance ratio at $z = 0.57$ |

[a] For dynamical dark energy models with constant equation of state, we denote the equation of state by $w$ and adopt the same prior as for $w_0$.

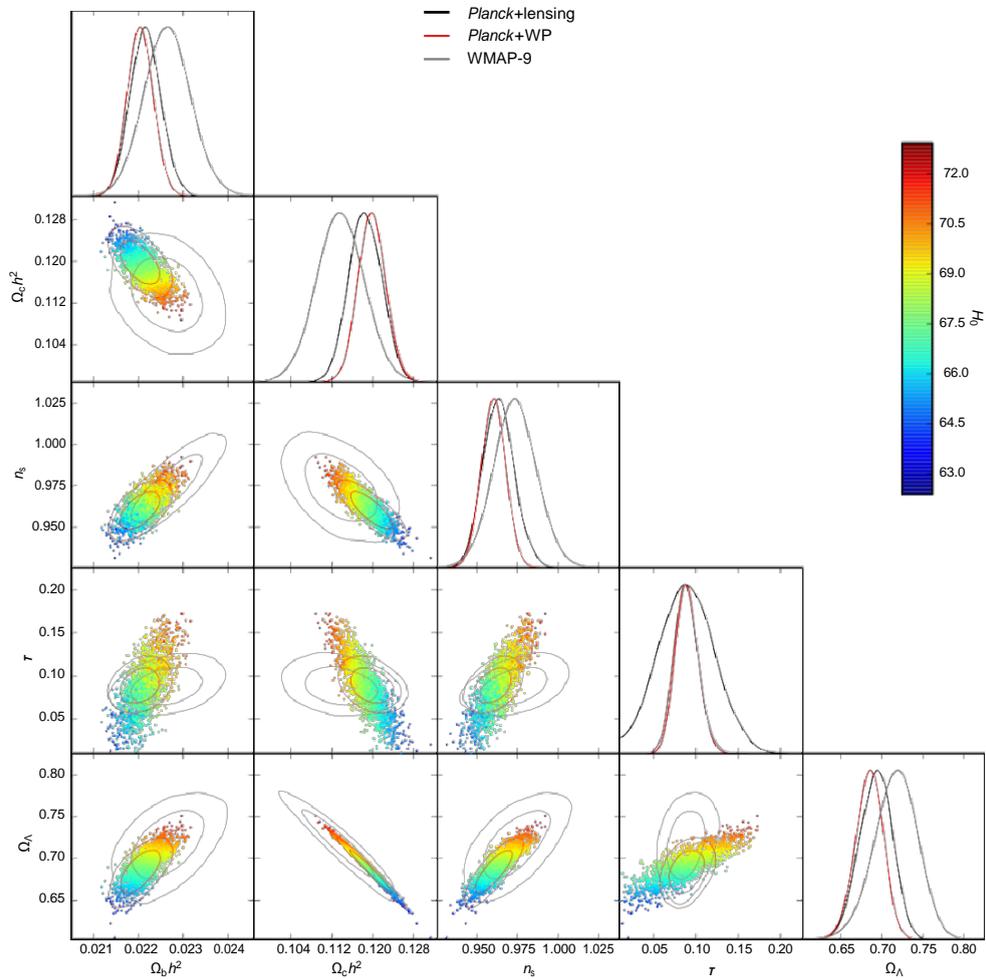

**Fig. 0.26** Comparison of the base ΛCDM model parameters for *Planck*+lensing only (colour-coded samples), and the 68% and 95% constraint contours adding WMAP low-$\ell$ polarization (WP; red contours), compared to WMAP-9 (Bennett *et al.* 2012*b*; grey contours).

---

[18] Annexes of the *Planck* parameter paper analysed the consistency of *Planck* versus SPT and *WMAP*, to verify that these data could be used jointly. We did in particular a detailed analysis of the (several) sources of the shift away from the parameters of the *WMAP*-7+SPT combination.

**Table 0.3** Cosmological parameter values for the *Planck*-only best-fit 6-parameter ΛCDM model (*Planck* temperature data plus lensing) and for the *Planck* best-fit cosmology including external data sets (*Planck* temperature data, lensing, *WMAP* polarization [WP] at low multipoles, high-$\ell$ experiments, and BAO, labelled [Planck+WP+highL+BAO]). The six parameters fit are above the line; those below are derived from the same model. Definitions and units for all parameters can be found in Table 0.2.

| Parameter | Planck (CMB+lensing) | | Planck+WP+highL+BAO | |
|---|---|---|---|---|
| | Best fit | 68 % limits | Best fit | 68 % limits |
| $\Omega_b h^2$ | 0.022242 | 0.02217 ± 0.00033 | 0.022161 | 0.02214 ± 0.00024 |
| $\Omega_c h^2$ | 0.11805 | 0.1186 ± 0.0031 | 0.11889 | 0.1187 ± 0.0017 |
| $100\theta_{MC}$ | 1.04150 | 1.04141 ± 0.00067 | 1.04148 | 1.04147 ± 0.00056 |
| $\tau$ | 0.0949 | 0.089 ± 0.032 | 0.0952 | 0.092 ± 0.013 |
| $n_s$ | 0.9675 | 0.9635 ± 0.0094 | 0.9611 | 0.9608 ± 0.0054 |
| $\ln(10^{10} A_s)$ | 3.098 | 3.085 ± 0.057 | 3.0973 | 3.091 ± 0.025 |
| $\Omega_\Lambda$ | 0.6964 | 0.693 ± 0.019 | 0.6914 | 0.692 ± 0.010 |
| $\sigma_8$ | 0.8285 | 0.823 ± 0.018 | 0.8288 | 0.826 ± 0.012 |
| $z_e$ | 11.45 | $10.8^{+3.1}_{-2.5}$ | 11.52 | 11.3 ± 1.1 |
| $H_0$ | 68.14 | 67.9 ± 1.5 | 67.77 | 67.80 ± 0.77 |
| Age/Gyr | 13.784 | 13.796 ± 0.058 | 13.7965 | 13.798 ± 0.037 |
| $100\theta_*$ | 1.04164 | 1.04156 ± 0.00066 | 1.04163 | 1.04162 ± 0.00056 |
| $r_{drag}$ | 147.74 | 147.70 ± 0.63 | 147.611 | 147.68 ± 0.45 |

Of course, the high-$\ell$ experiments do help in breaking the degeneracies between the foreground parameters which we have noted in a *Planck* alone analysis (with little impact on cosmology, since foreground and cosmological parameters are only weakly correlated, as we saw in fig. 0.21). And these experiments will help in strengthening constraints on extension to the base ΛCDM model when they entail modification of the highest-$\ell$ tail, say at $\ell \gtrsim 2000$. Therefore in the following, the reference CMB constraints will indeed come from the combination *Planck*+WP+high-$\ell$. This data combination yields in particular $n_s = 0.9585 \pm 0.0070$, now excluding scale invariance at the 5.9 σ level from the CMB alone.

I now turn to the confrontation of the prediction of *Planck* best fit base ΛCDM model with other observables (after reminding that the lensing potential spectrum was in reasonable agreement with the prediction based on this model, as well as $Y_P$). One such check is again internal, using the CMB polarisation from *Planck*. While preparing for the 2013 analysis and data release, we found that *Planck* polarised maps Q & U were failing to pass some of our null tests for consistency, in particular on the largest scales, demonstrating the existence of some residual systematics effects. This lead us to defer a full polarisation analysis to 2014, when we shall hopefully understand fully the origin of the effect and build a proper error model at all scales. But we nevertheless proceeded to make a simple polarisation analysis, *assuming* (as

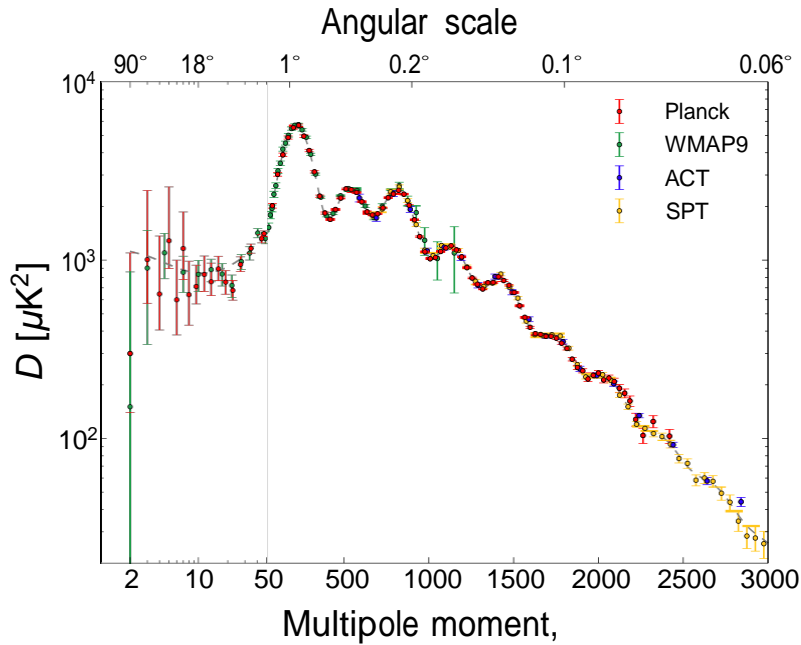

**Fig. 0.27** Measured angular power spectra of *Planck*, *WMAP*-9, ACT, and SPT. The model plotted is *Planck*'s best-fit model including *Planck* temperature, *WMAP* polarization, ACT, and SPT (or *Planck*+WP+HighL). Error bars include cosmic variance. The horizontal axis is logarithmic up to $\ell = 50$, and linear beyond.

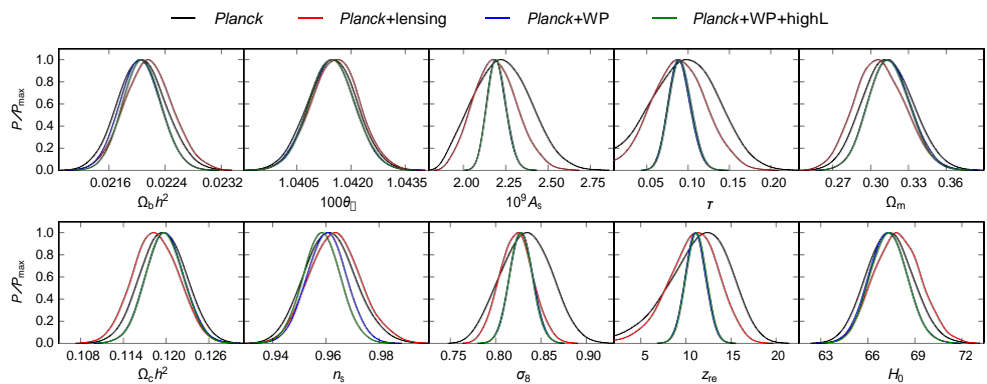

**Fig. 0.28** CMB constraints on the six $\Lambda$CDM parameters for various data combination.

our present understanding suggests) that the systematic effect found at large scale is simply negligible at smaller scales $\ell \gtrsim 50$. Figure 0.29-a) shows the resulting determination of the EE spectrum (blue points), as based on a simple analysis of the 143 and 217 GHz masked maps, together with the prediction (red line) of the *Planck* best fit model *as fitted on the temperature*

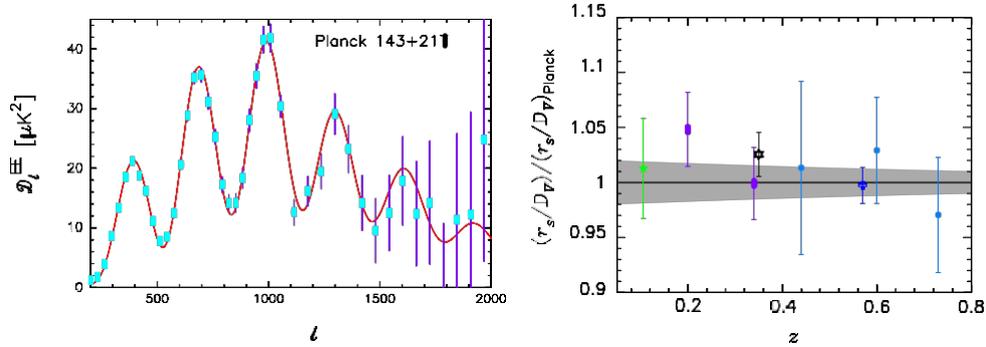

**Fig. 0.29** a) *Planck* EE spectrum. The red lines show the polarization spectra from the base ΛCDM *Planck*+WP+highL model, *which is fitted to the TT data only.* b) Acoustic-scale distance ratio $r_s/D_V(z)$ divided by the distance ratio of the same base ΛCDM model (P+WP+HL). The points are colour-coded as follows: green star (6dF); purple squares (SDSS DR7 as analyzed by Percival *et al.* 2010); black star (SDSS DR7 as analyzed by Padmanabhan *et al.* 2012); blue cross (BOSS DR9); and blue circles (WiggleZ). The grey band shows the approximate ±1 σ range allowed by *Planck*.

*data only.* The same good match holds for the TE spectrum. It is thus very comforting to see that *Planck* polarisation data is consistent with the model prediction based on the independent analysis of temperature anisotropies.

The low redshift acoustic scale, as determined by analysing baryon acoustic oscillations (BAO) in the power spectrum of the galaxies distribution, is primarily a geometric measurement of a well understood fundamental scale. It is generally recognised as less prone to subtle and complex gastrophysical effects than most astrophysical datasets. Figure 0.29-b) shows the acoustic distance ratio, $r_s/D_V$, expressed in units of the predicted ratio by the *Planck* CMB model (P+WP+HL, as for polarisation). The agreement is stunning (as opposed to the clear systematic offset with respect to the prediction of a fit based on WMAP7+SPT, see fig.B1 of *Planck* parameter paper). This lead us to use the fit based on *Planck*+WP+High-**f** + BAO as our reference "CMB+LSS" model, whose parameters values are given in Table 0.3, and which can be compared there to the *Planck* alone (CMB+lensing) ones. Maybe the most striking is that the simple base model still holds with an increase of the $n_s$ determination accuracy, $n_s = 0.9608 \pm 0.0054$; scale invariance in this model is now excluded at the 7.3 σ level.

Figure 0.30-a) shows the constraint set by *Planck* on $\Omega_m$ and $H_0$. Within the base ΛCDM model, the sound horizon ratio $\theta_*$ is proportional to the combination $\Omega_m h^3$, which is therefore very precisely determined (the transverse dimension of the ellipse is quite narrow). This tightly links low values of $\Omega_m$ with high values of $H_0$ (and $n_s$) when moving away from the best fit in one direction along the line of lesser constraint. As already mentioned, the *Planck* best fit value for $H_0$ is relatively low, and Figure 0.30-b) offers a quantitative comparison with a number of previous determination. The *Planck* value appears quite consistent with the *WMAP*-9 value[19], but appears low as compared to the direct $H_0$ measurements by Riess *et al.* 2011 (*HST* observations of Cepheid variables in the host galaxies of eight SNe Ia) and Freedman *et al.* 2012 (*Carnegie Hubble Program*); the other points are from Freedman *et al.* 2001 for the

---

[19] Actually the SPT12+*WMAP*-9 value is virtually identical to the SPT12+(*Planck*-restricted to **f** < 800).

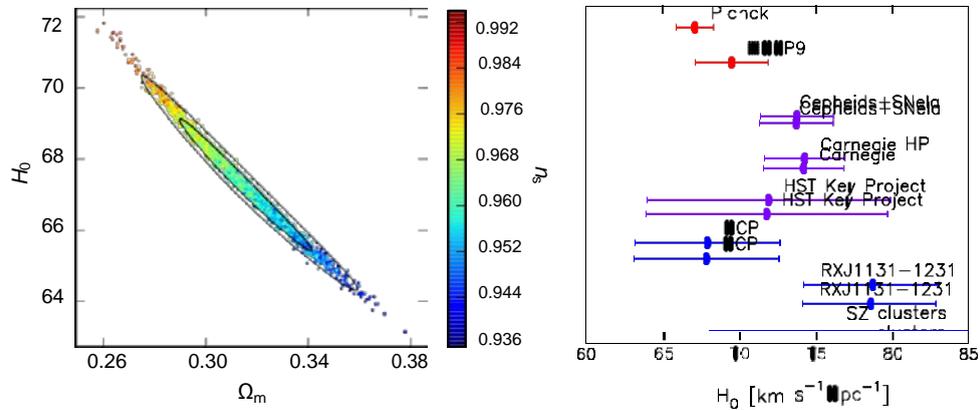

**Fig. 0.30** a) The sound horizon constraint, and the residual degeneracy between $\Omega_m$, $H_0$, $n_s$. Points show samples from the *Planck*-only posterior, coloured by the corresponding value of the spectral index $n_s$. The contours (68% and 95%) show the improved constraint from *Planck*+lensing+WP. The degeneracy direction is significantly shortened by including WP, but the well-constrained direction of constant $\Omega_m h^3$ (set by the acoustic scale), is determined almost equally accurately from *Planck* alone. b) Comparison of $H_0$ measurements, with estimates of ±1 σ errors, from a number of techniques (see text), with the spatially-flat ΛCDM model constraints from *Planck* and *WMAP*-9.

*HST* Key Project; "MCP" is from the Megamaser Cosmology Project (Braatz *et al.*, 2013); the point labelled "RXJ1131-1231" derives from gravitational lensing time delay measurements (Suyu *et al.*, 2013); finally, the point labelled SZ clusters is from Bonamente *et al.* 2006.

Given this tension with the direct but difficult, low-$z$, determination of $H_0$, it is quite interesting then to look at what distance measurements to type Ia supernovae tell us since, in the base ΛCDM model, the SNe data provide a constraint on $\Omega_m$ independent of the CMB. We compared with two SNe Ia samples: the sample of 473 SNe as reprocessed by Conley *et al.* 2011, which we refer to as the "SNLS" compilation; and the updated Union2.1 compilation of 580 SNe described by (Suzuki *et al.*, 2012). While the values for the latter are on the low side but remain compatible with that of *Planck*, our analysis as well as the analyses of Conley *et al.* 2011 and Sullivan *et al.* 2011, shows that the SNLS combined compilation favours a lower value of $\Omega_m$ than we find from the CMB.

In summary, within the base ΛCDM context, the *Planck* CMB fit is completely compatible with all BAO data, and shows some tension with some of the measurements of $H_0$ and $\Omega_m$ from low-$z$ distance measurements. It is worth mentioning that since the *Planck* data and preprints release on March 21$^{st}$, there have been a number of developments that affect some of the constraints from the supplementary astrophysical data just mentioned. Indeed, Humphreys *et al.* 2013 presented the final results of a long-term campaign to establish a new geometric maser distance to NGC4258. Their revised distance leads to a lowering of the Hubble constant, based on the Cepheid distance scale, partially alleviating the tension between the Riess *et al.* 2011 results and the *Planck* results on $H_0$. And in a recent paper, Betoule *et al.* 2013 present results on an extensive programme that improves the photometric calibrations of the SDSS and SNLS supernovae surveys. After this photometric recalibration the SNLS analysis of ΛCDM (Betoule *et al.*, 2014) favours a higher value of $\Omega_m$, consistent with the *Planck* base ΛCDM results[20]. Both shifts then go in the direction of narrowing the initial gap with *Planck* analysis. Let us wait and see what further data and analyses will lead us to.

#### 0.4.2 Extensions to ΛCDM

In this subsection we have a quick look at the post-*Planck* 2013 status of common extensions to the base ΛCDM model. Results regarding single parameter extensions are summarised in Table 0.4. This table also exhibits the tightening of the constraints from *Planck*+WP alone when either BAO, or high-**f** CMB information, or both are added. The short summary is that we found no compelling indication for any such extension. Still, these absences of detection are quite informative and are further commented below; when not specified, the numbers quoted in the text below are 95% confidence limits arising from the joint constraints from *Planck*+WP+High-**f**+BAO.

**Table 0.4** Constraints on one-parameter extensions to the base ΛCDM model. Data combinations all include *Planck* combined with *WMAP* polarization, and results are shown for combinations with high-**f** CMB data and BAO. Note that we quote 95% limits here.

| | Planck+WP | | Planck+WP+BAO | | Planck+WP+highL | | Planck+WP+highL+BAO | |
|---|---|---|---|---|---|---|---|---|
| Parameter | Best fit | 95% limits | Best fit | 95% limits | Best fit | 95% limits | Best fit | 95% limits |
| $\Omega_K$ | −0.0105 | $-0.037^{+0.043}_{-0.049}$ | 0.0000 | $0.0000^{+0.0066}_{-0.0067}$ | −0.0111 | $-0.042^{+0.043}_{-0.048}$ | 0.0009 | $-0.0005^{+0.0065}_{-0.0066}$ |
| $w$ | −1.20 | $-1.49^{+0.65}_{-0.57}$ | −1.076 | $-1.13^{+0.24}_{-0.25}$ | −1.20 | $-1.51^{+0.62}_{-0.53}$ | −1.109 | $-1.13^{+0.23}_{-0.25}$ |
| $\Sigma m_\nu$ [eV] | 0.022 | < 0.933 | 0.002 | < 0.247 | 0.023 | < 0.663 | 0.000 | < 0.230 |
| $N_{\text{eff}}$ | 3.08 | $3.51^{+0.80}_{-0.74}$ | 3.08 | $3.40^{+0.59}_{-0.57}$ | 3.23 | $3.36^{+0.68}_{-0.64}$ | 3.22 | $3.30^{+0.54}_{-0.51}$ |
| $Y_P$ | 0.2583 | $0.283^{+0.045}_{-0.048}$ | 0.2736 | $0.283^{+0.043}_{-0.045}$ | 0.2612 | $0.266^{+0.040}_{-0.042}$ | 0.2615 | $0.267^{+0.038}_{-0.040}$ |
| $dn_s/d\ln k$ | −0.0090 | $-0.013^{+0.018}_{-0.018}$ | −0.0102 | $-0.013^{+0.018}_{-0.018}$ | −0.0106 | $-0.015^{+0.017}_{-0.017}$ | −0.0103 | $-0.014^{+0.016}_{-0.017}$ |
| $r_{0.002}$ | 0.000 | < 0.120 | 0.000 | < 0.122 | 0.000 | < 0.108 | 0.000 | < 0.111 |

We first note that curvature is tightly constrained to zero, with $\Omega_{\text{tot}} = 1$ to within ~ 0.7%. It is worth remarking that now the CMB data alone, thanks to the precise determination of the lensing effect (which smooths the peak and troughs of the temperature spectrum), breaks the $\Omega_m - \Omega_\Lambda$ degeneracy and already gives by itself $100\,\Omega_K = -0.7 \pm 1.0$ (improving to −0.07 ± 0.33 when BAO are included).

Since the CMB probes the low-$z$ universe only through lensing, the inclusion of BAO brings nearly a factor of two improvement on possible deviation from -1 of the dark energy equation of state parameter $w$, which still remains indistinguishable (within 25%) from a cosmological constant; the constraints on the two parameter extension $(w_0, w_a)$ as in $w(a) \equiv p/\rho = w_0 + (1-a)w_a$ are of course even weaker.

The constraints on the sum of the neutrino masses are more interesting. Assuming for instance 3 active neutrinos of mass $m_\nu = \sum m_\nu/3 > 0.06$ eV (and $N_{\text{eff}} = 3.046$, *i.e.*, no additional $\nu$-like relativistic particles at decoupling), we found $\sum m_\nu < 0.23$ eV (95%CL CMB+BAO). The constraint from *Planck* arises mostly through the effect of neutrino masses via lensing (by $\ell = 1000$ the lensing potential is suppressed by $\sim 10\%$ in power for $m_\nu = 0.66$ eV); removing that source of information in the temperature spectrum[21] weakens the limit $m_\nu < 0.66$ eV (95%CL PT+WP+HL) which becomes $m_\nu < 1.08$ eV (95%CL PT+WP+HL). It is also worth noting that the (4-pt based) lensing likelihood would prefer higher values for $m_\nu$ (*i.e.*, this weakens the constraints). We also noted that the SZ number counts suggest a lower value of the small scale fluctuation today than that inferred from the CMB (this is measured by the $\sigma_8$ parameter which gives the linear amplitude of rms fluctuations today at the $8h^{-1}$ Mpc scale). If this tension is not due to an as yet unaccounted for gastrophysical effect in cluster formation (affecting the calibration of the SZ-based mass proxy), this could be accounted for by the effect of a larger $m_\nu$, somewhat above the 95% upper limit from CMB + BAO.

---

[20] Remarkably, the shift in contours in the $w-\Omega_m$ plane from the intersecting constraints of *WMAP*-9+"SNLS-old" to *Planck*+"SNLS-new" appear horizontal, *i.e.*, $w = -1$ is still preferred, but now for a higher $\Omega_m$ value.

On a somewhat related front, we found no evidence for additional neutrino-like relativistic particles beyond the three families of neutrinos in the standard model, as measured by possible deviation from the standard $N_{\text{eff}} = 3.046$ relativistic degrees of freedom at decoupling in the base model. Indeed we found the 95% CL of $3.30^{+0.54}_{-0.51}$ (CMB+BAO), with therefore no indication of extra massless relics. This therefore does not seem to confirm the hint of $3.84 \pm 0.4$ from *WMAP*-9+eCMB+BAO+$H_0$ which excluded 3 neutrinos families at more than $2\sigma$ (Bennett *et al.* 2012*b*; note though that including direct $H_0$ constraints would push $N_{\text{eff}}$ higher in *Planck*). It is also worth noting that the joint constraint on $N_{\text{eff}}$ and $m_\nu$ ($N_{\text{eff}} = 3.32^{+0.54}_{-0.52}$ $m_\nu < 0.28$ eV) does not differ much from the bounds obtained when introducing these parameters separately, *i.e.*, this degeneracy is now broken.

Leaving the helium abundance, $Y_P$, as a free parameter, leads to the 95%CL of $0.267^{+0.038}_{-0.040}$ which is of course very much weaker than the derived constraint if standard BBN is assumed ($Y_P = 0.2477 \pm 0.0001$). Let me also mention that we found no evidence of time variation of the fine structure constant[22] $\alpha/\alpha_0 = 0.9989 \pm 0.0037$ (*WMAP*-9 was $1.008 \pm 0.020$).

Regarding the constraints on initial conditions, it is convenient to expand the power spectra of primordial curvature and tensor perturbations on super-Hubble scales as

$$P_R(k) = A_s \left(\frac{k}{k_*}\right)^{n_s - 1 + \frac{1}{2} dn_s/d\ln k \ln(k/k_*) + \ldots}, \qquad (0.15)$$

$$P_t(k) = A_t \left(\frac{k}{k_*}\right)^{n_t + \frac{1}{2} dn_t/d\ln k \ln(k/k_*) + \ldots}, \qquad (0.16)$$

where $A_s$ ($A_t$) is the scalar (tensor) amplitude and $n_s$ ($n_t$), and $dn_s/d\ln k$ ($dn_t/d\ln k$) are the scalar (tensor) spectral index, and the running of the scalar (tensor) spectral index, respectively. We did not find any indication of running of the power law spectral index of the scalar perturbations, $dn_s/d\ln k = -0.014^{+0.016}_{-0.017}$, but these uncertainties are anyway much larger than predictions in simple models of inflation. Much more interesting are the constraint on $r_{0.002}$, the tensor to scalar ratio of primordial power at the pivot scale $k_0 = 0.002$ Mpc$^{-1}$. We found a 95% CL of $r_{0.002} < 0.11$ (CMB+BAO), which gives interesting constraints on the possible energy scale of inflation ($V_* = (1.94 \times 10^{16} \text{ GeV})^4 (r/0.12)$). Figure 0.31 compares our constraints in the $n_s - r$ plane with predictions from a number of common inflation models, showing that concave potentials are preferred (with both $\dot\varphi^2$ and $\dot\varphi^2/\rho$ increasing during single-field slow-roll inflation). Other lectures in the series will expand on the implications of these all important constraints. One may also refer to the "Encyclopaedia Inflationaris" paper (Martin *et al.*, 2013) to detail in only ~ 400 pages the constraints on the specific parameters of all single field inflation models.

---

[21] This is done by marginalising over an additional parameter, $A_L$, the amplitude of the lensing effect in units of the expectation in the standard model, which is normally set to unity.

[22] A different fine structure constant, $\alpha$, would shift the energy levels and binding energy of hydrogen and helium, inducing a modification of the ionization history of the Universe.

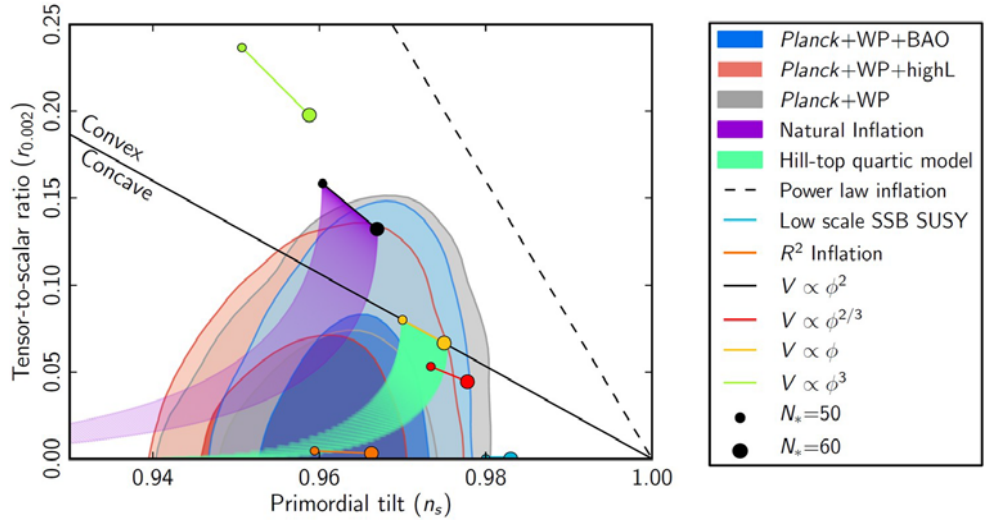

**Fig. 0.31** Marginalized 68% and 95% confidence levels for $n_s$ (the scalar spectral index of primordial fluctuations) and $r_{0.002}$ (the tensor to scalar power ratio at the pivot scale $k = 0.002\,\mathrm{Mpc}^{-1}$) from *Planck*+WP, alone and combined with high-$\ell$ and BAO data, compared to the theoretical predictions of selected inflationary models.

Let me simply mention here some of the main analyses and results of the paper we devoted to some further constraints on inflationary models based on two-point statistics (Planck Collaboration XXII, 2013). First we provided the constraints when both $r_{0.002}$ and other extensions ($N_{\rm eff}$, $Y_{\rm P}$, $w$, $m_\nu$, a generalised reionization) were allowed to vary for various data combination. The main point there is that introducing $N_{\rm eff}$ or $Y_{\rm P}$ shift the preferred value of $n_s$ towards somewhat higher values (by less than 0.02), and broadens the uncertainty contours, now weakening the exclusion of the $n_s = 1$ case to only two sigma. Since considerable uncertainty surrounds what occurred from the end of inflation till the end of entropy generation, we explored a number of scenarios to see hot the $n_s - r$ constraints can be somewhat modified (but with $n_s = 1$ still squarely excluded) and provided specific constraints on the additional effective equation of state parameter during that intermediate phase. We also searched for features in several ways.

First we searched for features by performing a reconstruction of the primordial scalar spectrum or of the best fitting inflaton potential. Figure 0.32-a) shows the deviation of the reconstructed power spectrum from a power law of index $n_s = 0.9603$, which shows no evidence of significant deviation. This plot was obtained for a roughness penalty $\lambda = 10^5$. This parameter controls a regularisation term added to the likelihood in this context, *i.e.*, when it is used to describe the primordial power as power in $\gtrsim 10$ bins rather than by a parametrised

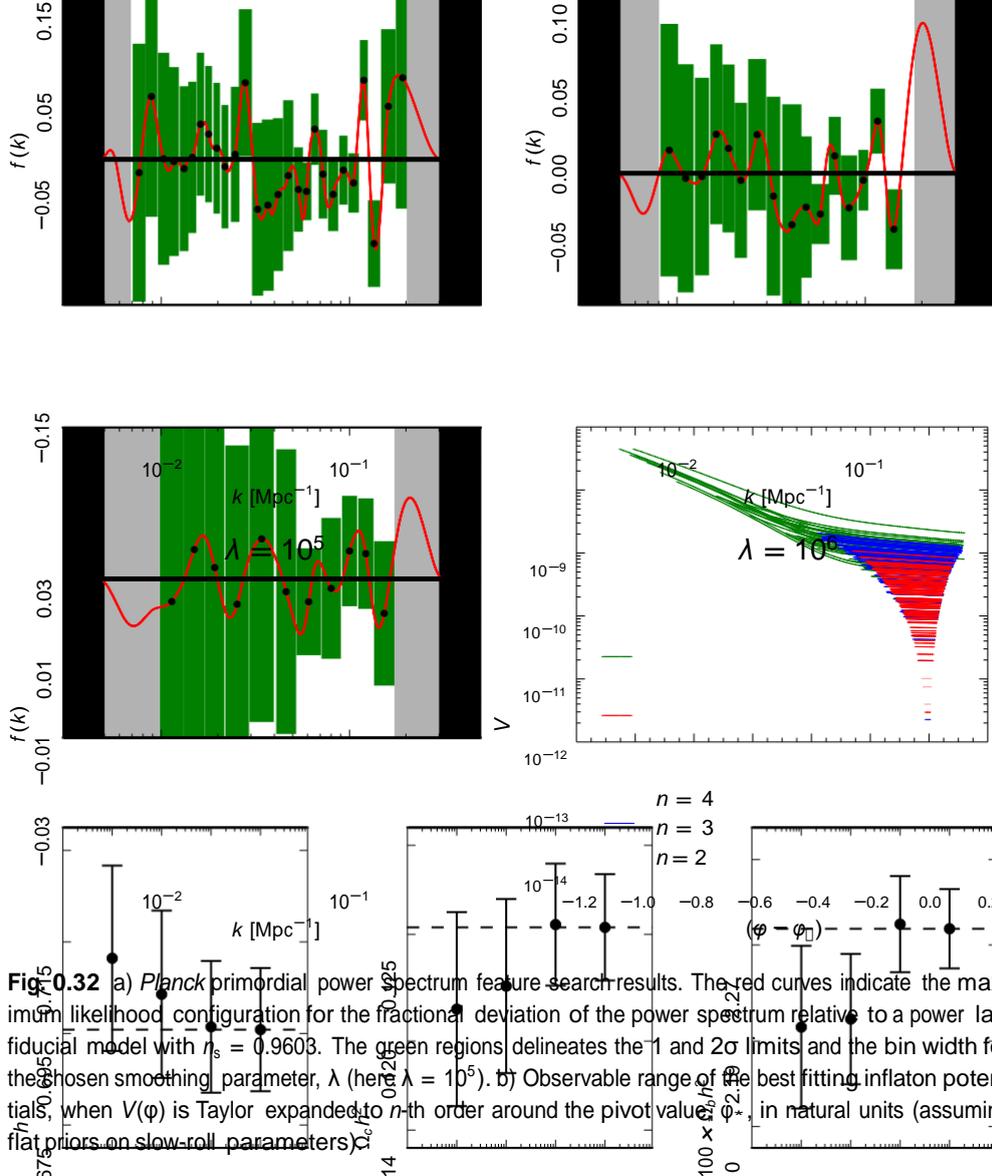

**Fig. 0.32** a) *Planck* primordial power spectrum feature search results. The red curves indicate the maximum likelihood configuration for the fractional deviation of the power spectrum relative to a power law fiducial model with $n_s = 0.9603$. The green regions delineates the 1 and $2\sigma$ limits and the bin width for the chosen smoothing parameter, $\lambda$ (here $\lambda = 10^5$). b) Observable range of the best fitting inflaton potentials, when $V(\varphi)$ is Taylor expanded to $n$-th order around the pivot value, $\varphi_*$, in natural units (assuming flat priors on slow-roll parameters).

function with a smaller number of degrees of freedom (e.g. the usual $A_s$, $n_s$). Increasing $\lambda$ would further smooth the curve and reduce the number of bins while increasing their width. With a smaller roughness penalty, that is, for $\lambda = 10^4$ or $\lambda = 10^3$ (and 20-25 bins), we were puzzled by a nominally statistically significant feature, clearly visible around $k = 0.13\,\mathrm{Mpc}^{-1}$. In our March draft, we cautioned that we did not account yet for the "look elsewhere" in the significance. But in any case, as already mentioned above, we since found that the power spectra, in particular at 217 GHz include a weak residual systematics around $\ell = 1800$ which is due to a non-perfect removal of the effect of an electromagnetic interference line arising from a non-nominal drive electronics of the 4K cryogenic stage. Since a feature at $k = 0.13\,\mathrm{Mpc}^{-1}$ induces one dip precisely at $\ell = 1800$, at least part of the significance comes from that. The draft of the article has been updated accordingly.

Figure 0.32-b) shows another reconstruction, that of the best fitting inflaton potential within the observable range, $V(\varphi)$, when $V(\varphi)$ is Taylor expanded to $n$-th order around the pivot value, $\varphi_*$, and the evolution is followed numerically. The figure shows samples extracted randomly from the converged Markov chains. It shows that the *Planck* data suggest a flat potential when the lowest order slow-roll primordial spectra are considered, but when this restriction is relaxed ($n = 4$, in green), the inflaton potential can differ markedly from a plateau-like potential, with a long and steep tail for $\varphi < \varphi_*$, with a kink around $\varphi \sim \varphi_* - 0.4$, which generates a significant running on the largest observable scales, while preserving a smaller running on smaller scales. With such a feature in the scalar primordial spectrum at large scales, combined with a non-zero contribution from tensor fluctuations, the best fit model for $n = 4$ has a temperature spectrum very close to that of the minimal $\Lambda$CDM model for $\ell \gtrsim 40$, but not for smaller multipoles. The large-scale data points from Planck are indeed low by $\sim 10\%$ as compared to the expectation from the ΛCDM model[23], see fig. 0.23. This kink in the potential allows the low-$\ell$ part to be fitted slightly better, but this is not unique. Another way to improve the low-$\ell$ fit is for instance to allow a small amount of isocurvature perturbations.

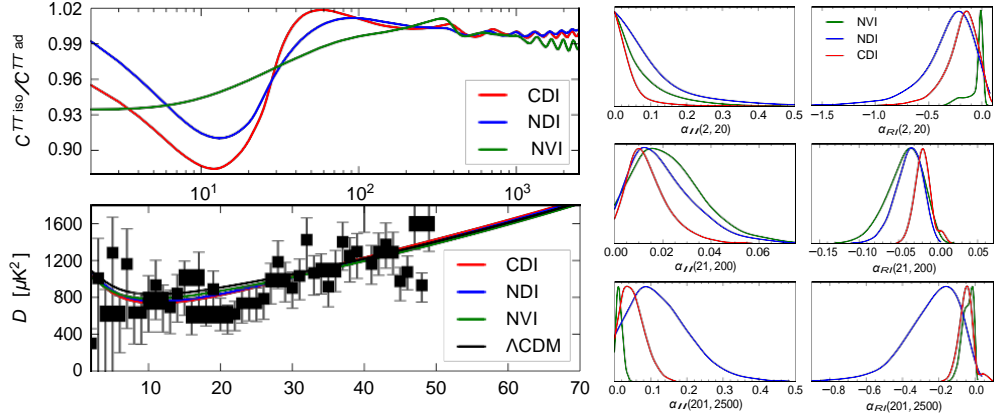

**Fig. 0.33** Contribution of isocurvature modes to the *Planck* +WP power spectrum (see text).

Indeed, fig. 0.33-a) gives the relative shape of various types of isocurvature modes divided by the adiabatic spectrum (top panel), and the best fit model at low-$\ell$ (below), using the *Planck* +WP power spectrum. In this analysis, we considered only one type of isocurvature modes at a time, either CDM isocurvature density (CDI), neutrino density (NDI), or neutrino velocity (NVI) isocurvature modes. The figure shows that, apart form the different phase of the oscillations, the main difference of these isocurvature modes $C_\ell$ shape relative to that of the standard adiabatic mode is the smaller amount of power at low $\ell$, which allows a better fit of the small deficit of power a $\ell \lesssim 40$ of the data versus that model. More quantitatively, fig. 0.33-b) gives the constraints on the fractional contribution, $\alpha_{XY}$ of isocurvature modes to the *Planck* +WP power spectrum, where $X$ and $Y$ can be either I for isocurvature or R for curvature perturbations:

$$\alpha_{XY}(\ell_{min}, \ell_{max}) = \frac{(\Delta T)^2_{XY}(\ell_{min}, \ell_{max})}{(\Delta T)^2_{tot}(\ell_{min}, \ell_{max})}, \quad \text{where } (\Delta T)^2_{XY}(\ell_{min}, \ell_{max}) = \sum_{\ell_{min}}^{\ell_{max}} (2\ell + 1) C_\ell^{X,Y}. \quad (0.17)$$

restricting the $\ell$ range allows to see what drives the fit. In any case, like for the less constrained shape of the inflation potential, the fit improvement is rather small. Let us mention here that the low-$\ell$ deficit also explains the fit improvement obtained with other model extensions, but it remains to be seen whether this deficit is a mere statistical fluctuations or an actual hint of new physics, which other probes like polarization might help reveal.

In the same paper on inflation, (Planck Collaboration XXII, 2013), we considered as well three models describing possible features in the primordial power spectrum, adding a global oscillation, a localized oscillation, or a cut-off to the large-scale power spectrum. These so-called "wiggles", "step-inflation" and "cut-off" model have more degrees of freedom and their inclusion improves the quality of the fit as compared to a pure power-law. Nevertheless, these models are not predictive enough, given our choice of prior, and the base $\Lambda$CDM model remains preferred. In any case, if real, these features are likely to show up in polarisation and Non-Gaussianity searches.

---

[23] This finding does not come from a revision of the low-$\ell$ $C_\ell$ points from WMAP, which are quite similar to those of *Planck*. It is rather than the additional high-$\ell$ information from *Planck* pins down quite accurately the parameters of the base ΛCDM model, which reduces the low-$\ell$ uncertainty of the model and reveals more strongly this tension.

We have also looked at constraints on inflationary models based on possible non-gaussianity, by using statistics of order higher than two (Planck Collaboration XXIV, 2013). First, fig. 0.34-a) offers one of the visualization we used of the 3-point function in harmonic space, *i.e.*, the bi-spectrum of temperature anisotropies at **f** low enough that we are not completely dominated by noise, which can for instance be used for graphically comparing with predictions of various models. This information can then be further condensed once a specific parametric model of the primordial bi-spectrum as been specified. We have used optimal and other estimators, (KSW, Binned, modal, skew-Cl, wavelets, Minkowski functionals...), since these different weighting might be affected differently by residual systematic effects. We have extensively tested for the latter using simulations and the 4 component separation methods of *Planck*, testing for the effect of masks and **f**$_{max}$ of the analysis, and debiased them when needed, in particular to account for the ISW-lensing part or residual compact sources.

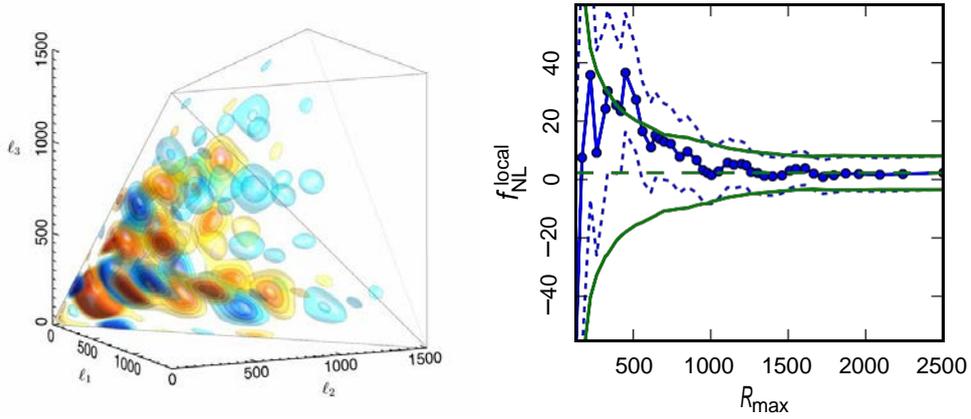

**Fig. 0.34** a) b).

Figure 0.34-b) shows the $f_{NL}^{local}$ constraints versus **f**$_{max}$ which nicely illustrates the fact that *Planck* does recover the relatively high value of WMAP9 at low resolution ($f_{NL}^{local}$ − *WMAP* = 37.2 ± 19.9), *i.e.*, for **f**$_{max}$ ~ 500, before settling down to zero with tighter uncertainties when more mores are included. *Planck* constraints stop improving at **f** ≳ 1500 when noise becomes dominant. *Planck* therefore constrains quite tightly the non-gaussianities of local, equilateral, or orthogonal type:

$$f_{NL}^{local} = 2.7 \pm 5.8, \quad f_{NL}^{equil} = -42 \pm 75, \quad f_{NL}^{ortho} = -25 \pm 39. \tag{0.18}$$

We also derived constraints on early Universe scenarios that generate primordial non-Gaussianity, including general single-field models of inflation, excited initial states (non-Bunch-Davies vacua), and directionally-dependent vector models. We provided an initial survey of scale-dependent feature and resonance models. These results bound both general single-field and multi-field model parameter ranges, such as the speed of sound, $c_s \geq 0.02$ (95% CL), in an effective field theory parametrization, and the curvaton decay fraction $r_D \geq 0.15$ (95% CL). The Planck data significantly limit the viable parameter space of the ekpyrotic/cyclic scenarios (see A. Linde's contribution in this volume). The amplitude of the four-point function in the local model $\tau_{NL} < 2800$ (95% CL). Taken together, these constraints represent the highest precision tests to date of physical mechanisms for the origin of cosmic structure.

For completeness, note that we also derived *Planck* constraints on the mass per unit length of cosmic strings ($G\mu/c^2 < 1.5 \times 10^{-7}$ for Nambu strings and $G\mu/c^2 < 3.2 \times 10^{-7}$ for field theory strings) and other topological defects(Planck Collaboration XXV, 2013) and the geometry and topology of the Universe (Planck Collaboration XXVI, 2013). Of course, we also looked at generic non-Gaussianity (Planck Collaboration XXIII, 2013), confirming previous results from COBE and WMAP, but making quite unlikely that these large-scale anomalies would be be linked to instrumental systematics or foreground residuals.

Let me conclude by mentioning the really cool detection of the deformation of the sky pattern due to our local velocity with respect to the rest frame defined by the CMB photons. This Doppler effect has both an aberration and a modulation effect of the anisotropies (Planck Collaboration XXVII, 2013). There are effects of order $10^{-3}$, applied to fluctuations which are already one part in roughly $10^5$, so they are quite small. Nevertheless, it becomes detectable with the all-sky coverage, high angular resolution, and low noise levels of the *Planck* satellite. Of course this lead to a less precise determination of our velocity relative to the rest frame of the CMB than the dipolar effect ($v = 384$ km s$^{-1}$ ± 78 km s$^{-1}$ (stat.) ± 115 km s$^{-1}$ (syst.)).

## 0.5 Conclusions

*Planck* has achieved an unprecedented experimental combination of sensitivity, angular resolution, and frequency coverage. In March 2013, ESA and the *Planck* Collaboration released the initial cosmology products based on the the first 15.5 months of Planck data, along with a set of scientific and technical papers and a web-based explanatory supplement. These data are a treasure trove for astrophysics at large, and cosmology in particular.

We found that a base $\Lambda$CDM model with 6 parameters is a very good fit to the *Planck* temperature spectrum, with parameters ($n_s$, $\Omega_b$, $\Omega_c$, $\theta$) accurately determined by *Planck* alone, with the remaining ($A_s$, $\tau$) degeneracy alleviated by adding the large scale polarization constraint from WMAP (WP). The base model is fully consistent with two other Planck observables, the CMB lensing and polarization power spectra. *Planck* alone exclude scale invariance ($n_s = 1$) at more than 4$\sigma$. The base model is also fully consistent with Baryon Acoustic Oscillation (BAO) measurements. The situation regarding $\Omega_m$ from Supernovae is currently changing but appears promising, while there is some tension with direct $H_0$ determination.

Using the CMB probes from *Planck*+WMAP Polarization+ ACT & SPT in combination with BAO constraints, scale invariance is now excluded at ~ 7$\sigma$, and we find no compelling evidence for any additional parameter. *Planck* data was also searched for traces of primordial non-Gaussianities, here again with no compelling positive indication. The single field slow-roll inflation class of model therefore survived the most stringent test of Gaussianity performed to date, which of course limits less minimal alternatives.

*Planck* also found some "anomalies" for the base model, either firming up earlier detection or pointing to new ones, which makes the work towards the next release even more exciting. This next release should be before the end of 2014, and it will include twice more data and polarization.

## Acknowledgements

All results are those of the *Planck* collaboration. All further mistakes are mine :-).

The development of *Planck* has been supported by: ESA; CNES and CNRS/INSU-IN2P3-INP (France); ASI, CNR, and INAF (Italy); NASA and DoE (USA); STFC and UKSA (UK); CSIC, MICINN, JA and RES (Spain); Tekes, AoF and CSC (Finland); DLR and MPG (Germany); CSA (Canada); DTU Space (Denmark); SER/SSO (Switzerland); RCN (Norway); SFI (Ireland); FCT/MCTES (Portugal); and PRACE (EU). A description of the Planck Collaboration and a list of its members, including the technical or scientific activities in which they have been involved, can be found at http://www.sciops.esa.int/index.php?project=planck&page=Planck_Collaboration.